\documentclass[lettersize,journal]{IEEEtran}

\usepackage{amsmath,amsfonts}
\usepackage{algorithmic}
\usepackage{algorithm}
\usepackage{array}
\usepackage[caption=false,font=normalsize,labelfont=sf,textfont=sf]{subfig}
\usepackage{textcomp}
\usepackage{stfloats}
\usepackage{url}
\usepackage{verbatim}
\usepackage{graphicx}
\usepackage{cite}
\usepackage[printonlyused]{acronym}
\newacro{ris} [RIS] {reconfigurable intelligent surface}
\newacro{leo} [LEO] {low Earth orbit}
\newacro{fim} [FIM] {Fisher information matrix}
\newacro{mimo} [MIMO] {multiple-input multiple-output}
\newacro{los} [LoS] {line-of-sight}
\newacro{nlos} [NLoS] {non-line-of-sight}
\newacro{aod} [AoD] {angle-of-departure}
\newacro{aoa} [AoA] {angle-of-arrival}
\newacro{ue} [UE] {user equipment}
\newacro{ofdm} [OFDM] {orthogonal frequency-division multiplexing}
\newacro{rfc} [RFC] {radio-frequency chain}
\newacro{peb} [PEB] {position error bound}
\newacro{oeb} [OEB] {orientation error bound}
\newacro{veb} [VEB] {velocity error bound}
\newacro{awgn} [AWGN] {additive white Gaussian noise}
\newacro{ukf} [UKF] {unscented Kalman filter}
\newacro{3d} [3D] {three-dimensional}
\newacro{6d} [6D] {six-dimensional}
\newacro{9d} [9D] {nine-dimensional}
\newacro{rmse} [RMSE] {root mean squared error}
\newacro{crb} [CRB] {Cram\'er-Rao bound}
\newacro{cdf} [CDF] {cumulative distribution function}
\newtheorem{definition}{Definition}
\usepackage{color}

\usepackage{tikz} 
\usepackage[utf8]{inputenc}
\usepackage{pgfplots} 
\usepackage{tabu,longtable}
\usepackage{diagbox}
\usepackage{threeparttable}
\usepackage{makecell}
\usepackage{multirow} 
\usepackage{units} 
\usepackage{bm} 
\usepackage{amssymb} 
\newcommand{\TT}{\mathsf{T}}
\newcommand{\HH}{\mathsf{H}}

\newcommand{\av}{{\bf a}}
\newcommand{\bv}{{\bf b}}

\newcommand{\dv}{{\bf d}}

\newcommand{\fv}{{\bf f}}

\newcommand{\pv}{{\bf p}}

\newcommand{\wv}{{\bf w}}
\newcommand{\vv}{{\bf v}}
\newcommand{\xv}{{\bf x}}

\newcommand{\zv}{{\bf z}}

\newcommand{\Am}{{\bf A}}
\newcommand{\Bm}{{\bf B}}
\newcommand{\Cm}{{\bf C}}
\newcommand{\Dm}{{\bf D}}

\newcommand{\Hm}{{\bf H}}

\newcommand{\Jm}{{\bf J}}
\newcommand{\Km}{{\bf K}}

\newcommand{\Pm}{{\bf P}}
\newcommand{\Qm}{{\bf Q}}
\newcommand{\Rm}{{\bf R}}
\newcommand{\Sm}{{\bf S}}

\newcommand{\Xm}{{\bf X}}
\newcommand{\Ym}{{\bf Y}}
\newcommand{\Zm}{{\bf Z}}

\newcommand{\At}{{\rm A}}

\newcommand{\Dt}{{\rm D}}

\newcommand{\Rt}{{\rm R}}
\newcommand{\St}{{\rm S}}

\newcommand{\Ut}{{\rm U}}

\newcommand{\alphav}{\hbox{\boldmath$\alpha$}}

\newcommand{\nuv}{\hbox{\boldmath$\nu$}}

\newcommand{\zetav}{\hbox{\boldmath$\zeta$}}
\newcommand{\phiv}{\hbox{\boldmath$\phi$}}

\newcommand{\thetav}{\hbox{$\boldsymbol\theta$}}

\newcommand{\omegav}{\hbox{\boldmath$\omega$}}
\newcommand{\xiv}{\hbox{\boldmath$\xi$}}

\newcommand{\rhov}{\hbox{\boldmath$\rho$}}

\newcommand{\Gammam}{\hbox{\boldmath$\Gamma$}}

\newcommand{\Sigmam}{\hbox{\boldmath$\Sigma$}}

\newcommand{\Thetam}{\hbox{\boldmath$\Theta$}}
\newcommand{\Omegam}{\hbox{\boldmath$\Omega$}}

\begin{document}
\title{LEO- and RIS-Empowered User Tracking:\\ A Riemannian Manifold Approach}

\author{Pinjun~Zheng,~\IEEEmembership{Graduate Student Member,~IEEE},
Xing~Liu,~\IEEEmembership{Member,~IEEE},\\
and~Tareq~Y.~Al-Naffouri,~\IEEEmembership{Senior Member,~IEEE}

\thanks{This publication is based upon the work supported by the King Abdullah University of Science and Technology (KAUST) Office of Sponsored Research (OSR) under Award ORA-CRG2021-4695. The authors are with the Electrical and Computer Engineering Program, Division of Computer, Electrical and Mathematical Sciences and Engineering (CEMSE), King Abdullah University of Science and Technology (KAUST), Thuwal, 23955-6900, Kingdom of Saudi Arabia. (\emph{Corresponding author:~Xing~Liu})}
}

\markboth{This work has been submitted to the IEEE for possible publication.  Copyright may be transferred without notice.}{draft}
\maketitle

\begin{abstract}
Low Earth orbit~(LEO) satellites and reconfigurable intelligent surfaces~(RISs) have recently drawn significant attention as two transformative technologies, and the synergy between them emerges as a promising paradigm for providing cross-environment communication and positioning services. This paper investigates an integrated terrestrial and non-terrestrial wireless network that leverages LEO satellites and RISs to achieve simultaneous tracking of the three-dimensional~(3D) position, 3D velocity, and 3D orientation of user equipment~(UE). To address inherent challenges including nonlinear observation function, constrained UE state, and unknown observation statistics, we develop a Riemannian manifold-based unscented Kalman filter~(UKF) method. This method propagates statistics over nonlinear functions using generated sigma points and maintains state constraints through projection onto the defined manifold space. Additionally, by employing Fisher information matrices~(FIMs) of the sigma points, a belief assignment principle is proposed to approximate the unknown observation covariance matrix, thereby ensuring accurate measurement updates in the UKF procedure. Numerical results demonstrate a substantial enhancement in tracking accuracy facilitated by RIS integration, despite urban signal reception challenges from LEO satellites. In addition, extensive simulations underscore the superior performance of the proposed tracking method and FIM-based belief assignment over the adopted benchmarks. Furthermore, the robustness of the proposed UKF is verified across various uncertainty levels.
\end{abstract}


\section{Introduction}

In the realm of modern wireless systems, the importance of advanced user tracking mechanisms is increasingly prominent. Current tracking methodologies encompass a wide range, including global navigation satellite systems (GNSS)~\cite{hofmann2007gnss}, radio frequency identification (RFID)~\cite{7008512}, advanced computer vision techniques~\cite{manzanilla2019autonomous}, inertial navigation systems (INS)~\cite{grewal2007global}, acoustic sensors~\cite{5975255}, laser~\cite{4522195}, radar~\cite{bahl2000enhancements}, and wireless sensor networks~\cite{Talvitie2023Orientation}. Among these technologies, using wireless radio signals stands out as a particularly innovative and effective solution~\cite{8409950}. The integration of tracking with wireless communication systems not only addresses the localization needs but also enhances the efficiency and reliability of communication networks~\cite{Chen2022A}. Compared with other techniques, the high flexibility, expansive global reach, and inherent connectivity infrastructure of wireless communication systems make them well-suited for integration at different levels. In particular, the transition from the fifth generation (5G) to the sixth generation (6G) heralds an era characterized by ultra-fast speeds, minimal latency, and exceptional reliability in wireless networks~\cite{9349624,Dang2020Should}. This evolution significantly enhances the integration and adaptability of tracking systems across diverse domains, ranging from indoor scenarios to outdoor environments, and from ground-level applications to space-based contexts~\cite{9390169,9628162}.

The recent resurgence of \ac{leo} satellites, particularly in tandem with 5G and 6G networks, marks a significant evolution in the field~\cite{9502642,10399870}. LEO satellites provide global coverage and reduced latency, key to enhancing communication and real-time tracking, navigation, and location-based services. Recent research has been actively investigating the potential of LEO communication satellites for opportunistic navigation as viable supplements or even alternatives to the traditional satellite-based localization systems, i.e., GNSS~\cite{10388052,10140066}. This burgeoning interest is underpinned by a suite of distinctive advantages that LEO satellites offer over GNSS, including stronger signal power, vast constellations, and a broad spectrum of frequency diversity~\cite{Su2019Broadband,9800136}. In an era where precision, speed, and reliability are increasingly critical in both tracking and communication domains, the unique capabilities of LEO satellites present a compelling solution to meet these demands. 

Similar to GNSS, LEO satellite signals inevitably undergo significant attenuation and disruption when navigating through urban and indoor environments, primarily due to obstacles like building structures, multipath interference, and overall reduced signal strength~\cite{Zheng20235G}. To tackle these challenges, integrating LEO satellites with the terrestrial network emerges as a promising solution. \Acp{ris}~\cite{Liu2021Reconfigurable,Wymeersch2020Radio,Wang2024Wideband}, as innovative devices capable of intelligently adapting and enhancing incoming electromagnetic waves, can be strategically deployed on the ground to enhance the reception of LEO satellite signals. Numerous literature has highlighted the substantial potential of RISs in enhancing the positioning capabilities of advanced wireless communication systems, e.g.,~\cite{Chen2024Multi,He2022Beyond,Zheng2023Misspecified,Zheng2023JrCUP}. Nowadays, initial research has shown that the collaborative design between LEO satellites and RISs can further extend the coverage and efficiency of LEO satellite signals~\cite{10365519,9539541}, thus boosting the positioning and tracking services~\cite{zheng2023leo}. The vast coverage of LEO constellations and the adaptive capabilities of RISs come together to form a more connected and monitored integrated wireless network, unlocking new potential for advanced tracking solutions.

While signals of LEO communication satellites and RISs potentially exhibit inherent compatibility~\cite{zheng2023leo}, seamlessly integrating these two systems into a cohesive unit still requires sophisticated signal processing techniques. This paper considers applying LEO satellites and RISs within wireless communication systems, a terrestrial and non-terrestrial integrated network, for comprehensive user tracking. Recognizing that traditional \ac{3d} or \ac{6d} tracking may fall short in contexts where grasping the full spectrum of motion dynamics is paramount, particularly in advanced contexts like automated vehicles and aircraft~\cite{stevens2015aircraft}, this work employs \ac{9d} tracking to offer an exhaustive perspective. The 9D tracking encompasses the dynamical estimation of the \ac{ue}'s \ac{3d} position, \ac{3d} velocity, and \ac{3d} orientation, simultaneously. Such a comprehensive approach is essential for the understanding and management of an object's movement, aligning with the intricate requirements of high-precision interaction, control, automation, and decision-making processes. 

Addressing the 9D tracking problem in an integrated LEO satellites and RISs system is challenging due to the following reasons:
\begin{itemize}
	\item \textit{The observation function is nonlinear}. Typically, in a \ac{mimo} communication system, the observation for the \ac{ue} state estimation is served by the geometric channel parameters obtained from the channel estimation process~\cite{Chen2022A}. Such channel parameters can contain Doppler shifts, \acp{aoa}, \acp{aod}, and channel delays. However, the mappings between these channel parameters and the unknown \ac{ue} state are highly nonlinear, increasing the difficulty of the state estimation.
	\item \textit{The unknown \ac{ue} state is constrained}. While the \ac{ue} position and velocity lie in Euclidean space, the \ac{ue} orientation, usually characterized by a rotation matrix,\footnote{The orientation can also be represented by,~e.g., Euler angles, which is a parametrization of minimal dimension without constraints. However, such a parametrization would result in the so-called singularity problem, which can significantly degrade the estimation performance in certain areas of the state space~\cite{HERTZBERG201357}. For this reason, we choose to use the rotation matrix representation in this work.} is constrained in the Lie group SO(3)~\cite{9197489}. Such a non-Euclidean topological structure poses significant challenges for conventional tracking algorithms based on Euclidean space.  Hence, advanced signal processing techniques are required to deal with this hybrid constraint in the joint 9D state estimation. 
	\item \textit{The statistics of the observation are unknown and time-varying}. Usually, the statistics of the observation are essential in designing estimation algorithms. Since the observation (i.e., geometric channel parameters) is obtained from channel estimation, the uncertainty in these estimated channel parameters depends on both the geometric configuration of the system and the performance of the adopted channel estimation algorithm. Even if we assume an efficient channel estimator is available, the estimation uncertainty still relies on the true UE state, which is unknown and time-varying. Therefore, how to approximate this uncertainty and assign an appropriate belief to the observation remains a critical issue.
\end{itemize}

Aiming to resolve the aforementioned problems, this paper proposes a novel Riemannian manifold-based \ac{ukf} method for the 9D user tracking in a hybrid LEO satellite-RIS system.
The main contributions of this work can be summarized as follows:
\begin{itemize}
\item We define an advanced 9D user tracking system incorporating LEO satellites and RISs. A cooperative downlink signal transmission framework is proposed to coordinate different satellites and collect geometric channel parameters effectively. By exploring the acquired channel parameters and the \ac{ue}'s motion dynamics, this system offers a comprehensive tracking solution, allowing simultaneous determination of the \ac{3d} position, \ac{3d} velocity, and \ac{3d} orientation of the \ac{ue}.
\item To address the defined 9D tracking problem, we develop a constrained \ac{ukf} algorithm on the Riemannian manifold. Specifically, the proposed algorithm addresses the key issues in the considered 9D tracking problem as follows.
	\begin{itemize}
		\item By generating and propagating the sigma points, the employed \ac{ukf} enables the approximation of the first and second-order moments of the unknown state following nonlinear functions through sample mean and variance calculations.
		\item By applying the Riemannian manifold theory within the \ac{ukf} framework, the estimation steps are streamlined within a curvature of high-dimensional space, which ensures that the tracking solutions naturally meet the desired constraints.
		\item By computing the weighted mean of the Fisher information matrices over the generated sigma points, the uncertainty in the observation can be well approximated. Hence, an appropriate belief assignment principle is proposed, guaranteeing the accuracy of the measurement update steps in the \ac{ukf}.
	\end{itemize} 
\item Extensive simulations are conducted to evaluate the proposed system and algorithm.
	\begin{itemize}
		\item The system is evaluated in a complex simulation environment, where the \ac{ue}'s trajectory goes through the rural, suburban, and urban scenarios with different scattering strengths and \ac{los} probabilities. The results indicate that the integration of RISs can significantly enhance the tracking performance and mitigate the performance degradation in urban regions, particularly for the \ac{ue} position estimation. Apart from that, simulation reveals a trade-off between the transmission overhead and the tracking performance. Moreover, the impact of different system parameters is widely tested, providing insights into the deployment of such collaborative tracking systems.
		\item The performance of the proposed tracking algorithm is thoroughly validated. The proposed belief assignment principle demonstrates a significant performance enhancement over identical belief, rivaling the effectiveness of employing the true covariance matrix of observations in certain regions. Benchmarked against the classical unconstrained \ac{ukf}, we affirm that the proposed Riemannian manifold-based \ac{ukf} effectively preserves the SO(3) constraint. This preservation leads to a notable enhancement in the \ac{ue} orientation estimation, consequently improving position and velocity estimation accuracy. Furthermore, extensive testing under various levels of state and observation uncertainty underscores the robustness of our algorithm.
	\end{itemize}
\end{itemize}

The rest of the paper is organized as follows: Section~\ref{sec:system_model} delineates the system model that integrates LEO satellites and RISs, detailing the geometry model, the signal model, and the transmission pipeline. Section~\ref{sec:Dynamics} sets up the dynamic state-space model of the \ac{ue}'s motion with the corresponding uncertainty representation. Based on the defined models, Section~\ref{sec:UKF_RieM} meticulously describes our proposed Riemannian manifold-based tracking method, elucidating its components, operational dynamics, and algorithm details.  The proposed system and algorithms are evaluated through a sequence of simulation tests in Section~\ref{sec:results}. Finally, our conclusions are presented in Section~\ref{sec:conclusion}.

\section{System Model}
\label{sec:system_model}

\begin{figure}[t]
  \centering
  \includegraphics[width=\linewidth]{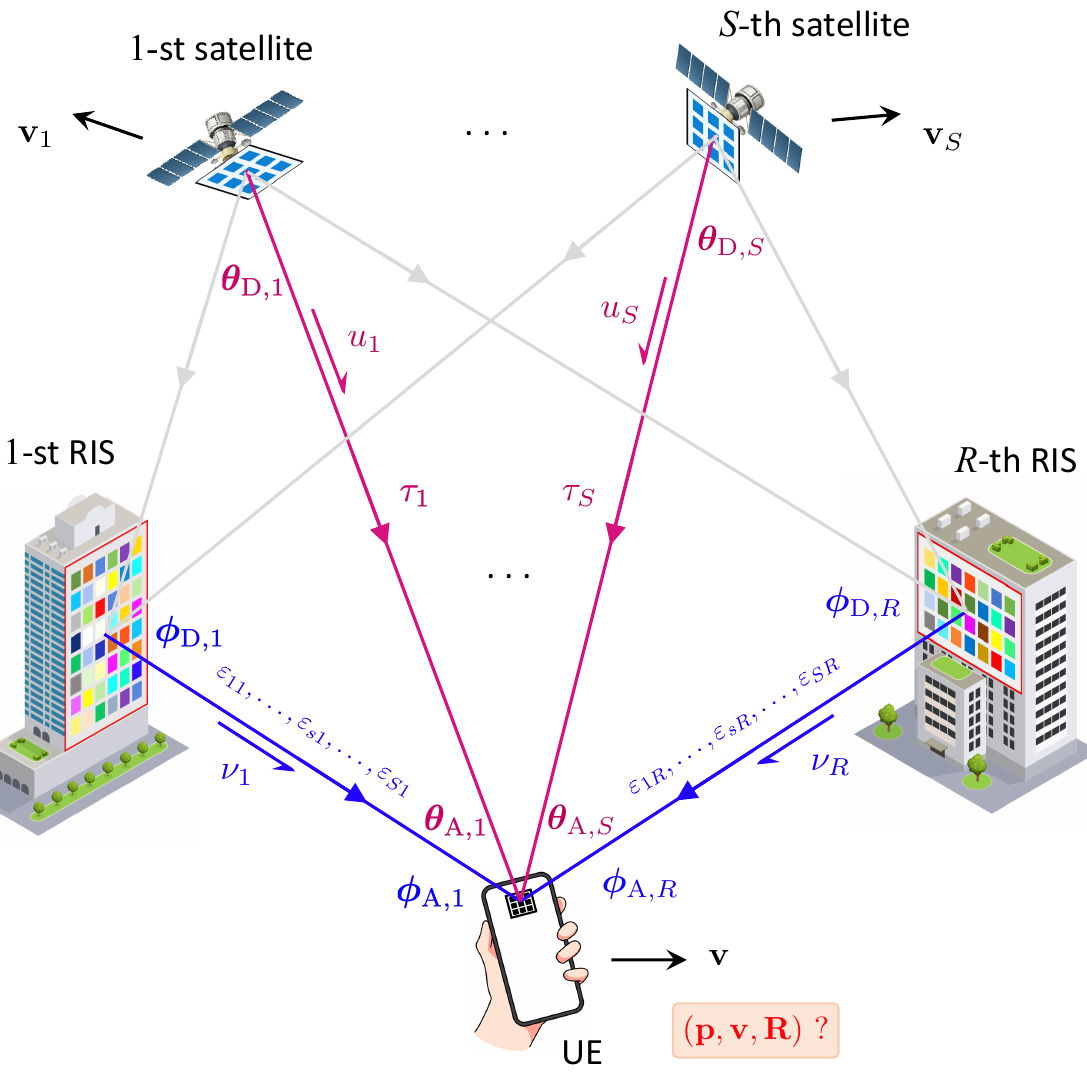}
  \caption{ 
      Illustration of the considered hybrid terrestrial-satellite system with~$S$ LEO satellites and~$R$ RISs. 
    }
  \label{fig_system}
\end{figure}

We consider a 9D user tracking problem in a hybrid terrestrial-satellite system consisting of~$S$ \ac{leo} satellites and~$R$ \acp{ris}, as depicted in Fig.~\ref{fig_system}. In this system, an unknown moving \ac{ue} receives downlink signals from both satellite-\ac{ue} direct channels and satellite-RIS-\ac{ue} reflection channels and then performs a 9D self-tracking, capturing the 3D position, the 3D velocity, and the 3D orientation of the \ac{ue} simultaneously. This section initially introduces the fundamental geometric relationship present in the system under consideration, followed by the signal model employed. Subsequently, we detail the timing framework for signal transmission and user tracking.

\subsection{Geometry Model}\label{sec_GM}
Within the proposed framework, the observation of the user tracking task is the geometric channel parameters, which are related to the user's position, velocity, and orientation through the geometric relationship. Therefore, this subsection presents the geometry model of the considered system and clarifies the underlying channel parameters, based on which the tracking algorithm will be developed further ahead. {It is noteworthy that during the tracking process, the position and velocity of LEO satellites can be obtained from the ephemerides contained within the Two-Line Element (TLE) files, which are updated daily by the North American Aerospace Defense Command (NORAD)~\cite{celestrak2022,10388052}. In addition, LEO satellites may also be equipped with a variety of onboard navigation systems such as GNSS receivers, inertial measurement units (IMUs), star trackers, and sun sensors, which provide accurate measurements of their position, orientation, and velocity~\cite{selvan2023precise,sarvi2020design}. On the other hand, RISs, which are assumed stationary in this study, can have their position and orientation precisely determined once deployed. As such, we assume the states of all the satellites and RISs are known when tracking the \ac{ue}.}

\begin{figure}[t]
  \centering
  \includegraphics[width=0.9\linewidth]{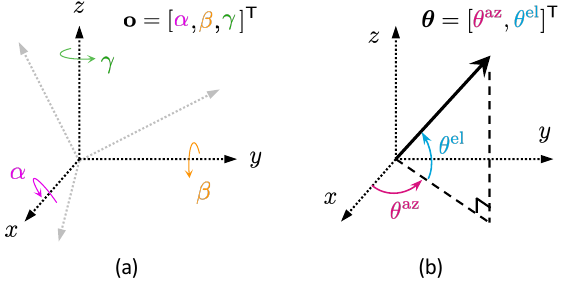}
  \caption{ 
      Illustration of the geometric setups related to coordinate systems. (a)~The definition of the Euler angles~$\alpha$, $\beta$, and~$\gamma$. (b)~The definition of the azimuth angle~$\theta^{\mathrm{az}}$ and the elevation angle~$\theta^{\mathrm{el}}$. 
    }
  \label{fig_angles}
\end{figure}

The known position and velocity of the $s$-th LEO satellite is denoted by~$\pv_{s}\in\mathbb{R}^3$ and~$\mathbf{v}_s\in\mathbb{R}^3$, respectively, while its orientation is represented by a rotation matrix~$\Rm_s\in\text{SO(3)}$. Here,~$\text{SO(3)}$ stands for the group of 3D rotations defined as 
\begin{equation}\label{eq:SO3}
	\text{SO(3)}\stackrel{\Delta}{=}\{\Rm|\Rm^\TT\Rm=\mathbf{I}_3,\mathrm{det}(\Rm)=1\}.
\end{equation} 
Specifically, the matrix~$\Rm_s$ describes the rotational connection between the body coordinate system of the~$s$-th satellite and the global coordinate system. For example, for arbitrary vector~$\dv$ at the global coordinate system, we can express it at the~$s$-th satellite's body coordinate system as~$\Rm_s^\TT\dv$.
Note that the degree of freedom of rotation matrices is~3. Such rotational relationships can be determined by 3D Euler angles~$\{\alpha,\beta,\gamma\}$, which are visualized in Fig.~\ref{fig_angles}-(a) and have the transform relation as~\cite{Liu2022Constrained,Chen2022A}
\begin{equation}
\Rm = \begin{bmatrix}
\mathcal{C}_\alpha\mathcal{C}_\beta & \mathcal{C}_\alpha\mathcal{S}_\beta\mathcal{S}_\gamma - \mathcal{S}_\alpha\mathcal{C}_\gamma & \mathcal{C}_\alpha\mathcal{S}_\beta\mathcal{C}_\gamma + \mathcal{S}_\alpha\mathcal{S}_\gamma\\
\mathcal{S}_\alpha\mathcal{C}_\beta & \mathcal{S}_\alpha\mathcal{S}_\beta\mathcal{S}_\gamma + \mathcal{C}_\alpha\mathcal{C}_\gamma & \mathcal{S}_\alpha\mathcal{S}_\beta\mathcal{C}_\gamma - \mathcal{C}_\alpha\mathcal{S}_\gamma\\
-\mathcal{S}_\beta & \mathcal{C}_\beta\mathcal{S}_\gamma & \mathcal{C}_\beta\mathcal{C}_\gamma
\end{bmatrix}.	
\end{equation}
Here,~$\mathcal{C}_\varphi$ and $\mathcal{S}_\varphi$ ($\varphi\in\{\alpha,\beta,\gamma\}$) are shorthand notations for $\cos{\varphi}$ and $\sin{\varphi}$, respectively.

The known position and orientation of the $r$-th \ac{ris} are denoted as~$\pv_r\in\mathbb{R}^3$ and~$\Rm_r\in\text{SO(3)}$, respectively. Similarly, we denote the unknown position, velocity, and orientation of the \ac{ue} as~$\pv\in\mathbb{R}^3$,~$\mathbf{v} \in\mathbb{R}^3$, and~$\Rm\in\text{SO(3)}$, respectively. Therefore, ~$(\pv,\mathbf{v},\Rm)$ represent the total unknown \ac{ue} state to be estimated. In this work, we assume the position and orientation of all RISs remain fixed over time, while the states of the \ac{ue} and satellites undergo dynamic changes. For notational convenience, we temporarily omit time indexes from these variables and reintroduce them starting from Section~\ref{sec:Dynamics}.

In this work, we assume downlink \ac{mimo} transmissions from \ac{leo} satellites to the \ac{ue} with the reflection through \acp{ris}.
Let~$M_s$,~$M_r$, and~$M$ denote the number of antennas at the~$s$-th LEO satellite, the~$r$-th RIS, and the \ac{ue}, respectively. Typically, several channel parameters related to the \ac{ue}'s state can be extracted through the channel estimation process based on the received signals from each of these \ac{mimo} channels. Therefore, we now define the involved channel parameters and clarify the geometric relationship between these channel parameters and the \ac{ue}'s state. Note that these channel parameters are associated with the corresponding \ac{los} paths.\footnote{In this work, we only utilize the channel parameters within the \ac{los} paths to perform user tracking, while the multipath effect is treated as additive noise.}

\subsubsection{The Satellite-\ac{ue} Direct Channel}
Taking the direct channel between the~$s$-th satellite and the \ac{ue} as an example, we investigate the channel parameters including the Doppler frequency shift, the \ac{aod} at the satellite, the \ac{aoa} at the \ac{ue}, and the channel delay. 

\textbf{The Doppler shift:} We denote the Doppler shift observed from the channel between the~$s$-th satellite and the \ac{ue} as~$u_s\in\mathbb{R}$, which can be defined as
\begin{equation} \label{eq_nu}
    u_s = \frac{ (\mathbf{v}_s - \mathbf{v} )^\TT(\pv -\pv_{s})}{\lambda\|\pv -\pv_{s}\|_2}.
\end{equation} 
Here,~$\lambda$ is the wavelength of the center frequency.

\textbf{The \ac{aod} at satellite:} The \ac{aod} at the $s$-th satellite is denoted by~$\thetav_{\Dt,s}\in\mathbb{R}^2$. Note that in 3D space, each \ac{aod} (or \ac{aoa}) here consists of an azimuth angle and an elevation angle. i.e.,~$\thetav_{\Dt,s}=[\theta_{\Dt,s}^{\mathrm{az}},\theta_{\Dt,s}^{\mathrm{el}}]^\TT$. The definition of the azimuth and elevation angles are presented in Fig.~\ref{fig_angles}-(b).
According to the underlying geometric relationship, these~\acp{aod} can be expressed in terms of the satellite and \ac{ue} states as
\begin{align}
    \theta_{\Dt,s}^{\mathrm{az}} &= \text{atan2}\big([\Rm_s^\TT(\pv -\pv_{s})]_2,[\Rm_s^\TT(\pv -\pv_{s})]_1\big), \label{eq_thetaDaz}\\
    \theta_{\Dt,s}^{\mathrm{el}} &= \text{asin}\big([\Rm_s^\TT(\pv -\pv_{s})]_3/\|\pv -\pv_{s}\|_2\big), \label{eq_thetaDel}
\end{align}
where~$[\cdot]_i$ indicates the $i$-th entry of a vector.

\textbf{The \ac{aoa} at \ac{ue}:} The \ac{aoa} observed at the \ac{ue} for the signals transmitted from the~$s$-th satellite is denoted by~$\thetav_{\At,s}=[\theta_{\At,s}^{\mathrm{az}},\theta_{\At,s}^{\mathrm{el}}]^\TT\in\mathbb{R}^2$. Analogously, these \acp{aoa} can be expressed as 
\begin{align}
    \theta_{\At,s}^{\mathrm{az}} &= \text{atan2}\big([\Rm^\TT(\pv_{s}-\pv)]_2,[\Rm^\TT(\pv_{s}-\pv)]_1\big), \label{eq_thetaAaz}\\
    \theta_{\At,s}^{\mathrm{el}} &= \text{asin}\big([\Rm^\TT(\pv_{s}-\pv)]_3/\|\pv_{s}-\pv\|_2\big). \label{eq_thetaAel}
\end{align}

\textbf{The channel delay:} In addition, we denote the time delay of the channel between the~$s$-th satellite and the \ac{ue} as~$\tau_s\in\mathbb{R}$, which can be expressed as
\begin{equation} \label{eq_tau}
   \tau_s = \frac{\|\pv_{s}-\pv \|_2}{c} + b_s,
\end{equation}
where $b_s$ is an unknown but fixed clock bias between the~$s$-th LEO satellite and the \ac{ue}.

\subsubsection{The Satellite-RIS-\ac{ue} Reflection Channel}
As demonstrated in Fig.~\ref{fig_system}, there are $S\times R$ different satellite-RIS-\ac{ue} channels. However, these reflection channels share the same~$R$ RIS-\ac{ue} subchannels. Moreover, since the states of both LEO satellites and RISs are known, the informative channel parameters related to the \ac{ue} state are concentrated in the~$R$ RIS-\ac{ue} subchannels only. Taking the subchannel between the~$r$-th RIS and the \ac{ue} as an example, we can similarly express the channel parameters in terms of the \ac{ris} and \ac{ue} states. 

\textbf{The Doppler shift:} The Dopper shift observed from the signals reflected through the~$r$-th RIS is denoted as~$\nu_r$, which can be defined as
\begin{equation}
	\nu_r = \frac{ \mathbf{v}^\TT(\pv_r-\pv )}{\lambda\|\pv_r-\pv\|_2}.
\end{equation}

\textbf{The \ac{aod} at RIS:} The \ac{aod} at the $r$-th RIS is denoted as~$\phiv_{\Dt,r}=[\phi_{\Dt,r}^\mathrm{az},\phi_{\Dt,r}^\mathrm{el}]^\TT\in\mathbb{R}^2$ and is given by
\begin{align}
    \phi_{\Dt,r}^{\mathrm{az}} &= \text{atan2}\big([\Rm_r^\TT(\pv \!-\!\pv_r)]_2,[\Rm_r^\TT(\pv \!-\!\pv_r)]_1\big), \label{eq_phiDaz}\\
    \phi_{\Dt,r}^{\mathrm{el}} &= \text{asin}\big([\Rm_r^\TT(\pv -\pv_r)]_3/\|\pv -\pv_r\|_2\big). \label{eq_phiDel}
\end{align}

\textbf{The \ac{aoa} at UE:} The \ac{aoa} observed at the \ac{ue} for the signals reflected through the~$r$-th RIS is denoted as~$\phiv_{\At,r}=[\phi_{\At,r}^{\mathrm{az}},\phi_{\At,r}^{\mathrm{el}}]^\TT\in\mathbb{R}^2$ and is given by 
\begin{align}
    \phi_{\At,r}^{\mathrm{az}} &= \text{atan2}\big([\Rm^\TT(\pv_r-\pv)]_2,[\Rm^\TT(\pv_r-\pv)]_1\big), \label{eq_thetaAaz}\\
    \phi_{\At,r}^{\mathrm{el}} &= \text{asin}\big([\Rm^\TT(\pv_r-\pv)]_3/\|\pv_r-\pv\|_2\big). \label{eq_thetaAel}
\end{align}

\textbf{The channel delay:} Finally, we define the time delay of the subchannel between the~$r$-th RIS and the \ac{ue}. Note that in these reflection channels, the signals transmitted from different satellites possess different channel delays. To distinguish this, we denote the time delay observed at the signals transmitted from the~$s$-the satellite and through the~$r$-th RIS as~$\varepsilon_{sr}$, which is defined as\footnote{Note that since the states of LEO satellites and RISs are known, the propagation delays over the satellite-RIS subchannels are known. Therefore, these parts of the delay are ignored in~$\varepsilon_{sr}$.}
\begin{equation} \label{eq_tau}
   \varepsilon_{sr} = \frac{\|\pv_r-\pv \|_2}{c} + b_s.
\end{equation}

\subsection{Signal Model}\label{sec_SigModel}

For each LEO satellite, we consider a downlink pilot transmission with~$K$ subcarriers. Assume all the LEO satellites and the \ac{ue} are equipped with a single \ac{rfc}. The received baseband signal at the \ac{ue} from the~$s$-th LEO satellite for the $k$-th subcarrier can be expressed as
\begin{equation}\label{eq:ysk}
	y_s^k = \sqrt{P_s}\wv_s^\HH\underline{\Hm}_s^k\fv_s x_s^k + n_s^k,
\end{equation}
where~$P_s\in\mathbb{R}$ denotes the average transmission power of the~$s$-th LEO satellite,~$\wv_s\in\mathbb{C}^{M }$ the combiner,~$\bar{\Hm}_s^k\in\mathbb{C}^{M \times M_\St}$ the wireless channel between the~$s$-th satellite and the \ac{ue},~$\fv_s\in\mathbb{C}^{M_\St}$ the precoder,~$x_s^k$ the transmitted unit-modulus symbol, and~$n_s^k\sim\mathcal{CN}(0,\sigma^2)$ the \ac{awgn}. Note that here we utilize analog precoders and combiners.

The wireless channel~$\bar{\Hm}_s^k$ can be decomposed as
\begin{equation}
	\underline{\Hm}_s^k = \Hm_{s}^k + \sum_{r=1}^R \Hm_{r}^k\Gammam_{r} \Hm_{sr}^k, 
\end{equation}
where~$\Hm_{s}^k\in\mathbb{C}^{M \times M_\St}$,~$\Hm_{sr}^k\in\mathbb{C}^{M_\Rt\times M_\St}$,~and $\Hm_{r}^k\in\mathbb{C}^{M \times M_\Rt}$ represent the channel between the~$s$-th satellite and the \ac{ue}, the channel between the~$s$-th satellite and the~$r$-th RIS, and the channel between the~$r$-th RIS and the \ac{ue}, respectively. Here,~$\Gammam_{r}\in\mathbb{C}^{M_\Rt\times M_\Rt}$ is the reconfigurable reflection matrix of the~$r$-th RIS, which is a diagonal matrix and assumed frequency-independent under the narrowband assumption. The diagonal entry~$\Gamma_{r}(i,i)$ stands for the reflection coefficient of the~$i$-th unit cell in the RIS. Typically, a passive RIS keeps~$|\Gamma_{r}(i,i)|\leq 1$, while an active RIS can provide an extra amplitude gain so that~$|\Gamma_{r}(i,i)|> 1$, $\forall\ i=1,\dots,M_\Rt$. 

{
We consider a typical scenario where both the \ac{los} and \ac{nlos} paths are present in the near-ground space. The shadowed Rician model can be utilized to capture the statistics of the mixed channels, which aligns well with measurements and has been widely adopted in satellite channel modeling~\cite{Abdi2003New,Li2022Downlink,Jung2022Performance,Kim2023Downlink,He2024Physical,You2020Massive}. The Rician model treats the \ac{nlos} component of the received signal as a zero-mean Gaussian random variable while the \ac{los} component is deterministic. 
	Specifically, we can express the Satellite-UE channel~$\Hm_s^k$ and the RIS-UE channel~$\Hm_r^k$ as~\cite{Li2022Downlink,Jin2007On,Zheng2023Coverage}
	\begin{align}\label{eq:Rician}
		\Hm_\chi^k &= \sqrt{\frac{K_f}{K_f+1}}\bar{\Hm}_\chi^k + \sqrt{\frac{1}{K_f+1}}\tilde{\Hm}_\chi^k,
	\end{align}
	where~$\chi\in\{s,r\}$,~$K_f$ is the Rician factor,\footnote{{Typically, the Rician factor is crucial for applications such as link budget calculations and adaptive modulation, and it can be estimated from the envelope of the received signal~\cite{Medawar2013Approximate,Giunta2018Estimation}. However, the proposed tracking method in this paper does not rely on the knowledge of the Rician factor.}}~$\bar{\Hm}_\chi^k$ denotes the deterministic \ac{los} channel, and~$\tilde{\Hm}_\chi^k$ denotes the Gaussian \ac{nlos} component. For the satellite-RIS channel, however, we assume the RISs are deployed at elevated positions that are far away from the near-ground clutter, and thus there is no \ac{nlos} path between the satellite and RIS. Hence, we have~$\Hm_{sr}^k=\bar{\Hm}_{sr}^k$.

The \ac{los} channels between the satellite, \ac{ris}, and \ac{ue} can be expressed as~\cite{Chen2022A,Li2022Downlink,He2024Physical,Zheng2023Coverage}
\begin{align}
	\bar{\Hm}_s^k &= \alpha_{s} e^{j2\pi(tu_s-k\Delta f\tau_s)}\av (\thetav_{\At,s})\av_s^\TT(\thetav_{\Dt,s}),\label{eq:Hs}\\
	\bar{\Hm}_r^k &= \alpha_{r}e^{j2\pi(t\nu_r-k\Delta f\varepsilon_{sr})} \av (\phiv_{\mathrm{A},r}) \av_r^\TT(\phiv_{\mathrm{D},r}),\\
	\bar{\Hm}_{sr}^k &= \alpha_{sr} e^{j2\pi(t\upsilon_{sr}-k\Delta f\varsigma_{sr})} \av(\bm{\varphi}_{\At,sr})\av_s^\TT(\bm{\varphi}_{\Dt,sr}).\label{eq:Hsr}
\end{align}
Here,~$\alpha_{s}$,~$\alpha_{r}$, and~$\alpha_{sr}$ denote the complex channel gains,~$t$ denotes the time elapsed since the first transmission from the~$s$-th satellite,~$\Delta f=B/K$ is the subcarrier spacing with~$B$ denoting the signal bandwidth, and~$\av $ and~$\av_s$ represent the array response vectors of the \ac{ue} and the~$s$-th LEO satellite, respectively. The detailed expressions of these array response vectors can be found in, e.g.,~\cite{You2020Massive,Chen2022A,Zheng2023Coverage}. In addition, the parameters~$\upsilon_{sr}$,~$\varsigma_{sr}$,~$\bm{\varphi}_{\At,sr}$, and~$\bm{\varphi}_{\Dt,sr}$ in~\eqref{eq:Hsr}, which respectively denote the Doppler shift, time delay, \ac{aoa}, and \ac{aod} of the satellite-RIS channel, are assumed to be known since the states of both the LEO satellites and RISs are known.

The \ac{nlos} channels~$\tilde{\Hm}_\chi^k\in\mathbb{C}^{M\times M_\chi}$,~$\chi\in\{s,r\}$, are described as~$\tilde{\Hm}_\chi^k=\Thetam^{1/2}\check{\Hm}_\chi^k\Thetam^{1/2}_\chi$, based on the Rician model~\cite{Jin2007On,Abdi2003New,Goldsmith2005Wireless}. Here,~$\Thetam\in\mathbb{C}^{M\times M}$ and~$\Thetam_\chi\in\mathbb{C}^{M_\chi\times M_\chi}$ are the correlation matrices at the antenna arrays of the \ac{ue} and satellite/RIS, respectively. Each entry of~$\check{\Hm}_\chi^k\in\mathbb{C}^{M\times M_\chi}$ is an independent and identically distributed (i.i.d.)~$\mathcal{CN}(0,1)$ random variable. 
}

\subsection{Transmission and Tracking Framework}\label{sec_TTF}

\begin{figure}[t]
  \centering
  \includegraphics[width=\linewidth]{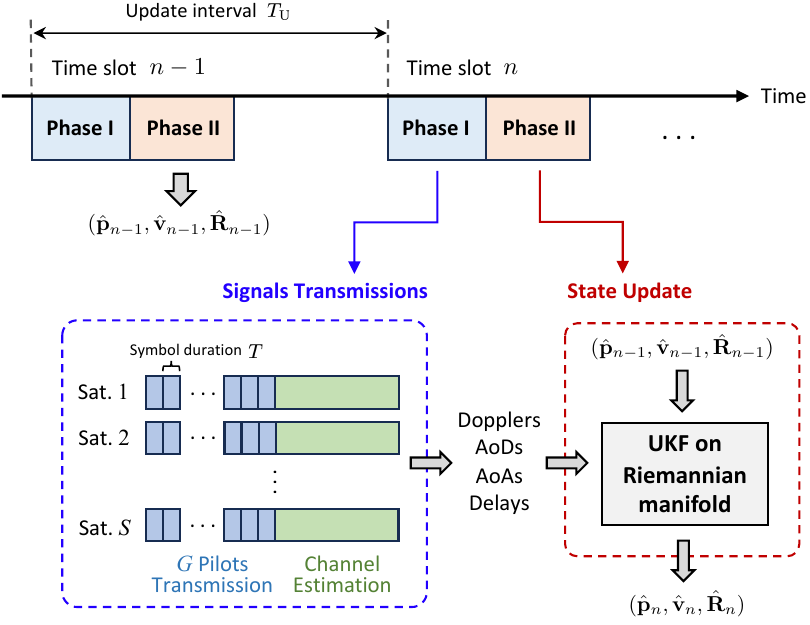}
  \caption{ 
      The frame structure of the proposed hybrid tracking system.  
    }
  \label{fig_framework}
\end{figure}

The temporal framework of the proposed tracking system is presented in Fig.~\ref{fig_framework}. Let~$T_\Ut$ denote the update interval of the moving \ac{ue}'s state estimation. We segment the tracking operation at each interval period into two sequential phases: Phase~I for signal transmission and Phase~II for state updates. 

In Phase~I, each of the LEO satellites first transmits~$G$ pilot symbols (each with symbol duration~$T$) to the \ac{ue}. For the~$g$-th transmission, the time~$t$ in~\eqref{eq:Hs}--\eqref{eq:Hsr} is given by
\begin{equation}
	t = (g-1)T,\quad g=1,2,\dots,G.
\end{equation}
 We assume that the~$G$ transmissions are completed within the channel coherence time, ensuring that the position, velocity, and orientation of the \ac{ue} remain constant. Furthermore, we adopt orthogonal frequency division for different satellites so that multiple satellites can transmit simultaneously.
After the pilot transmissions of each satellite, a channel estimation process is conducted to obtain the geometric channel parameters including Doppler shifts, \acp{aod}, \acp{aoa}, and channel delays, as elaborated in Section~\ref{sec_GM}. 

In Phase~II, we estimate and update the \ac{ue}'s 9D state based on i) the geometric channel parameters collected from Phase~I and ii) the \ac{ue} state estimate at the previous time slot. As a foundation to develop the methodology, Section~\ref{sec:Dynamics} comprehensively elucidates the dynamic state-space model governing the movement of the \ac{ue}. Subsequently, Section~\ref{sec:UKF_RieM} will showcase how the proposed \ac{ukf}, leveraging the Riemannian manifold, adeptly solves the state update problem under consideration.

\section{Dynamic State-Space Model}
\label{sec:Dynamics}
Based on the system model described in Section~\ref{sec:system_model}, the dynamic state-space model is presented in this section. We first clarify the time and measurement update relationships and then determine the underlying uncertainty representation.

\subsection{Discrete-Time State-Space Dynamics}

To formulate the tracking problem, we define the unknown state as
\begin{equation}\label{eq:zeta}
	\zetav = [\pv ^\TT,\vv ^\TT,\bv^\TT,\text{vec}(\Rm )^\TT]^\TT\in\mathbb{R}^{S+15}, 
\end{equation}
where~$\bv=[b_1,b_2,\dots,b_S]^\TT\in\mathbb{R}^S$ is the unknown clock bias between the satellites and \ac{ue}. 
The observations, on the other hand, are the geometric channel parameters defined in Section~\ref{sec_GM}, which we can concatenate as 
\begin{equation}\label{eq:eta}
	\rhov=[\rhov_0^\TT,\rhov_1^\TT,\dots,\rhov_S^\TT]^\TT,
\end{equation}
where 
\begin{align}
	&\rhov_0 = [\nuv^\TT,\phiv_\mathrm{D}^\TT,\phiv_\mathrm{A}^\TT]^\TT\in\mathbb{R}^{5R},\\
	&\rhov_s\! =\! \begin{bmatrix} u_s, \thetav_{\Dt,s}^\TT, \thetav_{\At,s}^\TT,\tau_s,\varepsilon_{s1},\!\dots,\!\varepsilon_{sr},\!\dots,\!\varepsilon_{sR}\end{bmatrix}^\TT\!\in\!\mathbb{R}^{R+6},\\
	&\nuv = \begin{bmatrix} \nu_{1},\dots,\nu_{r},\dots,\nu_{R} \end{bmatrix}^\TT\in\mathbb{R}^{R}, \\
	&\phiv_\mathrm{D} = \begin{bmatrix} \phiv_{\Dt,1}^\TT,\dots,\phiv_{\Dt,r}^\TT,\dots,\phiv_{\Dt,R}^\TT \end{bmatrix}^\TT\in\mathbb{R}^{2R}, \\
	&\phiv_\mathrm{A} = \begin{bmatrix} \phiv_{\At,1}^\TT,\dots,\phiv_{\At,r}^\TT,\dots,\phiv_{\At,R}^\TT \end{bmatrix}^\TT\in\mathbb{R}^{2R}.
\end{align}

Now we add the time indexes to the state vector~$\zetav$ and the observation vector~$\rhov$. Thus the observations across a series of discrete time instants can be denoted as~$\{\rhov_0,\rhov_1,\dots,\rhov_N\}$, and the corresponding states are denoted as~$\{\zetav_0,\zetav_1,\dots,\zetav_N\}$. Then, we can consider the following nonlinear dynamic system
\begin{align}
	&\textit{Time update:}\quad \zetav_{n+1} = f(\zetav_{n}),\label{eq:process}\\
	&\textit{Measurement update:}\quad \rhov_n = h(\zetav_n). \label{eq:observation}
\end{align}
Here,~$f(\cdot)$ is the process function and~$h(\cdot)$ is the observation function.
\subsubsection{Process Function}
The process function~$f(\cdot)$ is represented by the following mapping relationships:
\begin{equation}\label{eq:f}
	\begin{aligned}
	\pv_{n+1} &= \pv_{n} + \mathbf{v}_{n}\delta_t + \bar{\av}_{n}\frac{\delta_t^2}{2},\\
	\mathbf{v}_{n+1} &= \mathbf{v}_{n} + \bar{\av}_{n}\delta_t,\\
	\bv_{n+1} &= \bv_{n},\\
	\Rm_{n+1} &= \Rm_{n}\bar{\Omegam}_n,\\
\end{aligned}
\end{equation}
where~$\delta_t$ is the time interval between consecutive observations,~$\bar{\av}_{n}\in\mathbb{R}^3$ and~$\bar{\Omegam}_n\in\text{SO(3)}$ are the \ac{ue}'s acceleration and rotation at the $n$-th time instant. For clarification, we can rewrite~\eqref{eq:process} as~$\zetav_{n+1} = f(\zetav_{n};\bar{\av}_{n},\bar{\Omegam}_n)$. We assume that we can acquire a measurement of~$\bar{\av}_{n}$ and~$\bar{\Omegam}_n$ at each time instant through an inertial measurement unit~(IMU) in the \ac{ue}, which we can denote as
\begin{align}
	\hat{\av}_{n} &= \bar{\av}_{n} + \tilde{\av}_{n} \\
	\hat{\Omegam}_n &= \bar{\Omegam}_n\tilde{\Omegam}_n,
\end{align}
where~$\tilde{\av}_{n}\in\mathbb{R}^3$ and~$\tilde{\Omegam}_n\in\text{SO(3)}$ represent measurement errors that introduce perturbations on the state update.

\subsubsection{Observation Function}
The observation function maps the unknown \ac{ue} state to the observed channel parameters, which are represented by~\eqref{eq_nu}--\eqref{eq_tau}.
These channel parameters are, in turn, obtained by applying a channel estimation procedure and typically with certain estimation errors~\cite{He2021Channel,Chen2023Channel}. 

\subsection{Uncertainty Representation}
\label{sec:UR}

Before developing our tracking algorithm, it is essential to determine the statistics of the uncertainty in the time update and the measurement update. Specifically, we can rewrite~\eqref{eq:process} and~\eqref{eq:observation} as
\begin{align}
	&\textit{Time update:}\quad \zetav_{n+1} = f(\zetav_{n};\hat{\av}_{n},\hat{\Omegam}_n) + \Delta\zetav_{n+1},\label{eq:TUnew}\\
	&\textit{Measurement update:}\quad \hat{\rhov}_n = h(\zetav_n) + \Delta\rhov_n,\label{eq:MUnew}
\end{align}
where~$\Delta\zetav_{n+1}$ is the perturbation on the updated state introduced by the measurement errors~$\tilde{\av}_{n}$ and~$\tilde{\Omegam}_n$, and~$\Delta\rhov_n$ stands for the estimation error of the channel parameters~$\hat{\rhov}_n$.
Now we determine the statistics of~$\Delta\zetav_{n+1}$ and~$\Delta\rhov_n$, respectively. 

\subsubsection{Uncertainty on Time Update} 
\label{sec:UTU}
We assume that the error on the acceleration measurement is an~additive Gaussian noise~$\tilde{\av}_{n} = \hat{\av}_{n} - \bar{\av}_{n} \sim \mathcal{N}(\mathbf{0},\Cm_a)$. For the error on the rotation measurement, we model a small perturbation using a similar routine as~\cite{HERTZBERG201357}. First, we define a \ac{3d} Gaussian vector $\tilde{\omegav}_n\sim\mathcal{N}(\mathbf{0},\Cm_\omega)$. Then, the rotational perturbation~$\tilde{\Omegam}_n$ is modeled as a small rotation around axis~$\tilde{\omegav}_n\in\mathbb{R}^3$ with angle~$\|\tilde{\omegav}_n\|_2$. We will detail this operation in Section~\ref{sec:RMRS}. We assume that the covariance matrices of the IMU measurements,~i.e.,~$\Cm_a$ and~$\Cm_{\omega}$, are known.

Based on~\eqref{eq:f}, the perturbation source to each component of the state vector~$\zetav_{n+1}$, i.e.,~$(\pv_{n+1},\mathbf{v}_{n+1},\bv_{n+1},\Rm_{n+1})$, is~$(\tilde{\av}_{n}{\delta_t^2}/{2},\tilde{\av}_{n}\delta_t,\mathbf{0}_{S\times 1},\tilde{\omegav}_n)$.
When applying the classical \ac{ukf} without considering SO(3) constraint, we can assign a Gaussian belief to the state as
\begin{equation}
	\Delta\zetav_{n+1} \sim \mathcal{N}(\mathbf{0},\Sm),
\end{equation}
where~
\begin{equation}
	\Sm = \begin{bmatrix}
		\frac{\delta_t^4}{4}\Cm_a & \frac{\delta_t^3}{2}\Cm_a & \mathbf{0}_{3\times S} & \mathbf{0}_{3\times 9} \\
		\frac{\delta_t^3}{2}\Cm_a & \delta_t^2\Cm_a & \mathbf{0}_{3\times S} & \mathbf{0}_{3\times 9} \\
		\mathbf{0}_{S\times 3} & \mathbf{0}_{S\times 3} & \mathbf{0}_{S\times S} & \mathbf{0}_{S\times 9} \\
		\mathbf{0}_{9\times 3} & \mathbf{0}_{9\times 3} & \mathbf{0}_{9\times S} & \frac{\|\Cm_\omega\|_\mathrm{F}}{2\pi}\mathbf{I}_9
	\end{bmatrix}\in\mathbb{R}^{(S+15)\times(S+15)}.
\end{equation}
Note that this covariance matrix is an approximation that may cause performance loss since the perturbation on the \ac{ue} rotation is indeed not an additive variation. We will demonstrate that employing Riemannian manifold techniques can circumvent this approximation loss.

\subsubsection{Uncertainty on Measurement Update}

For the measurement update~\eqref{eq:MUnew}, the uncertainty~$\Delta\rhov_n$ comes from the estimation error of the channel parameters~$\rhov_n$. We assume that the required channel parameters are obtained through an efficient channel estimator. It then follows that the covariance matrix of~$\Delta\rhov_n$ is given by the inverse of the \ac{fim} of the underlying channel estimation problem~\cite{Kay1993Fundamentals,Zheng2023Coverage}.

To derive the \ac{fim}, let's rewrite the received signal~$y_s^k$ given in~\eqref{eq:ysk} as follows
\begin{equation}\label{eq_y}
	y_s^k = \ell_s^k + z_s^k + n_s^k,
\end{equation}
where~$\ell_s^k$ is the received signal propagating through the \ac{los} paths, and~$z_s^k$ contains all the \ac{nlos} multipath components.  
Based on the signal model in Section~\ref{sec_SigModel}, we have {
\begin{align}
	\ell_s^k &= \sqrt{\frac{K_fP_s}{K_f+1}}\wv_s^\HH\Big(\bar{\Hm}_s^k+\sum_{r=1}^R\bar{\Hm}_r^k\Gammam_r\Hm_{sr}^k\Big)\fv_sx_s^k,\\
	z_s^k &= \sqrt{\frac{P_s}{K_f+1}}\wv_s^\HH\Big(\tilde{\Hm}_s^k+\sum_{r=1}^R\tilde{\Hm}_r^k\Gammam_r\Hm_{sr}^k\Big)\fv_sx_s^k,
\end{align}
Since the entries of~$\check{\Hm}_s^k$ and~$\check{\Hm}_r^k$ are i.i.d. zero-mean unit-variance Gaussian random variables and~$n_s^k\sim\mathcal{CN}(0,\sigma^2)$, we can obtain that~
\begin{equation}
	z_s^k + n_s^k\sim\mathcal{CN}(0,C_s^k), \label{eq:newNoise}
\end{equation}
where
\begin{equation}
	C_s^k = \frac{P_s|x_s^k|^2\|\Thetam^{\frac{1}{2}}\wv_s\|_2^2\Big(\|\Thetam_s^{\frac{1}{2}}\fv_s\|_2^2 + \|\Thetam_r^{\frac{1}{2}}\Gammam_r\Hm_{sr}^k\fv_s\|_2^2\Big)}{K_f+1} + \sigma^2.\label{eq:Csk}
\end{equation}
}

Based on~\eqref{eq_y}--\eqref{eq:Csk}, the \ac{fim} of~$\rhov$, which is denoted as~$\Jm$, can be computed using the steps outlined in Appendix~\ref{appen:Jch}. Then we can write
\begin{equation}\label{eq:Jinv}
	\Delta\rhov_n\sim\mathcal{N}\big(\mathbf{0},\Sigmam_n\big),
\end{equation} 
where~$\Sigmam_n=\Jm^{-1}(\zetav_n)$.
Note that the computation of~$\Jm$ at time~$n$ requires the knowledge of the true \ac{ue} state~$\zetav_n$, which is typically not available in the tracking process. To show this dependency, here we further detail the notation of the \ac{fim} as~$\Jm(\zetav_n)$.

\section{The Proposed UKF on Riemannian Manifold}
\label{sec:UKF_RieM}
Since the observation function is nonlinear, the \ac{ukf} framework is chosen to solve the 9D tracking problem. The \ac{ukf} is a variation of the Kalman filter, bypassing the necessity for linearization and derivative computation by approximating the first and second-order moments of the state via a cluster of sigma points~\cite{Wan2000Nonlinear,Sayed2022Inference}. {As mentioned in Section~\ref{sec:UTU}, the classical \ac{ukf} cannot preserve the SO(3) constraint on the \ac{ue} orientation, which results in an inaccurate update of the state covariance matrix and degrades the estimation performance. To address this limitation, we employ the principles of Riemannian manifold theory. This theory offers a powerful toolset for navigating the smooth curvature of high-dimensional spaces and forces the states to fall into the constrained space. Combining the unique properties of the Riemannian manifold and the robust estimation capabilities of the \ac{ukf}, this section proposes a novel 9D tracking method grounded in these integrated frameworks. Section~~\ref{sec:RMRS} initially details the state representation and defines the necessary mathematical operations on Riemannian manifold. Following this, Section~~\ref{sec:RieMUKF} outlines the principal procedures of the UKF designed for implementation on Riemannian manifold.}

\subsection{State Representation via Riemannian Manifold Geometry}
\label{sec:RMRS}

By inspecting the state vector~$\zetav$ in~\eqref{eq:zeta}, we see the \ac{ue} orientation~$\Rm $ lies in the group of 3D rotation~SO(3), as defined in~\eqref{eq:SO3}. The group SO(3) is a smooth and parallelizable Riemannian submanifold of the Euclidean space~$\mathbb{R}^{3\times 3}$~\cite{9197489}.
On the contrary, the other variables in~$\zetav$ lie in the Euclidean space. Specifically, we have
\begin{equation}
	\pv \in\mathbb{R}^3,\ \mathbf{v} \in\mathbb{R}^3,\ \bv\in\mathbb{R}^S,\ \Rm \in\text{SO(3)}.
\end{equation}
Since the Cartesian product of two embedded submanifolds yields a new manifold, the state vector~$\zetav$ resides on the following compound manifold
\begin{equation}
	\zetav \in \mathbb{M} \triangleq \mathbb{R}^3 \times \mathbb{R}^3 \times \mathbb{R}^S \times \text{SO(3)}, 
\end{equation}
where~$\times$ represents the Cartesian product. According to the model in~\eqref{eq:f} and the analysis in Section~~\ref{sec:UTU}, for each updated point~$\zetav_{n+1}\in\mathbb{M}$, the corresponding perturbations are introduced by~$(\tilde{\av}_{n}{\delta_t^2}/{2},\tilde{\av}_{n}\delta_t,\mathbf{0}_{S\times 1},\tilde{\omegav}_n)\in\mathbb{R}^3\times\mathbb{R}^3\times\mathbb{R}^S\times\mathbb{R}^3$, {which can be re-expressed as~$[\tilde{\av}_{n}{\delta_t^2}/{2},\tilde{\av}_{n}\delta_t,\mathbf{0}_{S\times 1},\tilde{\omegav}_n]^\TT\in\mathbb{R}^{S+9}$, with statistics that follow
\begin{equation}
	[\tilde{\av}_{n}{\delta_t^2}/{2},\tilde{\av}_{n}\delta_t,\mathbf{0}_{S\times 1},\tilde{\omegav}_n]^\TT\sim\mathcal{N}(\mathbf{0},\Pm),
\end{equation}
where
\begin{equation}\label{eq:P}
	\Pm \!=\! \begin{bmatrix}
		\frac{\delta_t^4}{4}\Cm_a & \frac{\delta_t^3}{2}\Cm_a & \mathbf{0}_{3\times S} & \mathbf{0}_{3\times 3} \\
		\frac{\delta_t^3}{2}\Cm_a & \delta_t^2\Cm_a & \mathbf{0}_{3\times S} & \mathbf{0}_{3\times 3} \\
		\mathbf{0}_{S\times 3} & \mathbf{0}_{S\times 3} & \mathbf{0}_{S\times S} & \mathbf{0}_{S\times 3} \\
		\mathbf{0}_{3\times 3} & \mathbf{0}_{3\times 3} & \mathbf{0}_{3\times S} & \Cm_\omega
	\end{bmatrix}\in\mathbb{R}^{(S+9)\times(S+9)}.
\end{equation}
}

\begin{figure}[t]
  \centering
  \includegraphics[width=0.75\linewidth]{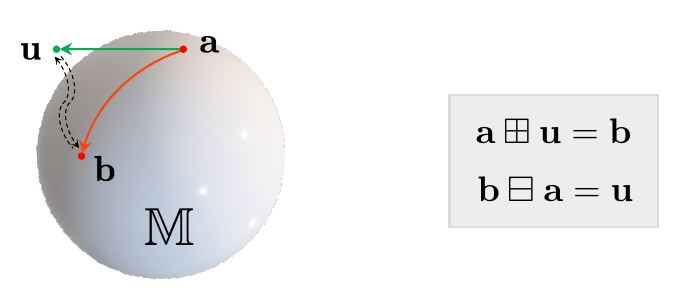}
  \caption{ 
      Visualization of the defined~$\boxplus$ and~$\boxminus$ operations.  
    }
  \label{fig_boxplus_minus}
\end{figure}

To accurately characterize the perturbation introduction and keep the SO(3) constraint, we define the following two encapsulation operators~\cite{HERTZBERG201357}, which are visualized in Fig.~\ref{fig_boxplus_minus}.

\begin{definition}[$\boxplus$~operation] \label{definition1}
For a manifold $\mathbb{F}$, the $\boxplus$ operation combines a point $\mathbf{a} \in \mathbb{F}$ and a Euclidean vector $\mathbf{u} \in \mathbb{R}^n$ to produce another point $\mathbf{b} = \mathbf{a} \boxplus \mathbf{u} \in \mathbb{F}$.
\end{definition}

\begin{definition}[$\boxminus$~operation] \label{definition2}
For two points $\mathbf{a}, \mathbf{b} \in \mathbb{F}$ on a manifold $\mathbb{F}$, and given a Euclidean vector $\mathbf{u} \in \mathbb{R}^n$ such that $\mathbf{b} = \mathbf{a} \boxplus \mathbf{u} \in \mathbb{F}$, the $\boxminus$ operation retrieves the Euclidean vector $\mathbf{u} = \mathbf{b} \boxminus \mathbf{a} \in \mathbb{R}^n$.
\end{definition}

Note that the defined operations~$\boxplus$ and~$\boxminus$ can be implemented in different ways. According to Definition~\ref{definition1}, for arbitrary~$(\pv ,\mathbf{v} ,\bv,\Rm )\in\mathbb{M}$ and~$(\mathbf{e}_1,\mathbf{e}_2,\mathbf{e}_3,\mathbf{e}_4)\in\mathbb{R}^3\times\mathbb{R}^3\times\mathbb{R}^S\times\mathbb{R}^3$, one realization of~$\boxplus$ is given by
\begin{multline}
	(\pv ,\mathbf{v} ,\bv,\Rm ) \boxplus (\mathbf{e}_1,\mathbf{e}_2,\mathbf{e}_3,\mathbf{e}_4) = (\pv +\mathbf{e}_1,\\ \mathbf{v} +\mathbf{e}_2,\bv+\mathbf{e}_3,\Rm \boxplus_{\text{SO(3)}}\mathbf{e}_4).
\end{multline}
Here,~$\Rm \boxplus_{\text{SO(3)}}\mathbf{e}_4$, as described in Section~\ref{sec:UTU}, denotes the operation that applies the rotation around axis~$\mathbf{e}_4\in\mathbb{R}^3$ with angle~$\|\mathbf{e}_4\|_2$ to~$\Rm $. Specifically, such operation can be conducted as~\cite{HERTZBERG201357},\cite[Sec.~4]{koks2006explorations}
\begin{equation}
	\Rm \boxplus_{\text{SO(3)}}\mathbf{e}_4 = \Rm \Rm_4,
\end{equation}
where
\begin{align}
	\Rm_4 &= \cos{\vartheta}\mathbf{I}_3 + (1-\cos{\vartheta})\av\av^\TT + \sin{\vartheta}\Bm, \\
	\vartheta &= \|\mathbf{e}_4\|,\quad \av = \mathbf{e}_4/\|\mathbf{e}_4\|,\\
	\Bm &= \begin{bmatrix}
		0 & -\mathbf{e}_4[3] & \mathbf{e}_4[2] \\
		\mathbf{e}_4[3] & 0 & -\mathbf{e}_4[1] \\
		-\mathbf{e}_4[2] & \mathbf{e}_4[1] & 0
	\end{bmatrix}.
\end{align}

According to Definition~\ref{definition2}, for arbitrary two points~$(\pv^a,\mathbf{v}^a,\bv^a,\Rm^a)$ and~$(\pv^b,\mathbf{v}^b,\bv^b,\Rm^b)$ in~$\mathbb{M}$, we can implement
\begin{multline}(\pv^a,\mathbf{v}^a,\bv^a,\Rm^a)\boxminus(\pv^b,\mathbf{v}^b,\bv^b,\Rm^b) = (\pv^a-\pv^b,\\ \mathbf{v}^a-\mathbf{v}^b,\bv^a-\bv^b,\Rm^a\boxminus_{\text{SO(3)}}\Rm^b).
\end{multline}
Here,~$\boxminus_{\text{SO(3)}}$ is the inverse operation of~$\boxplus_{\text{SO(3)}}$, which is given by~\cite{HERTZBERG201357}
\begin{align}
	&\Rm^a\boxminus_{\text{SO(3)}}\Rm^b = \frac{\varphi}{2\sin{\varphi}}\begin{bmatrix}
		D_{3,2} - D_{2,3}\\
		D_{1,3} - D_{3,1}\\
		D_{2,1} - D_{1,2}
	\end{bmatrix},\\
	&\varphi = \arccos{\frac{\text{Tr}(\Dm)-1}{2}},\ \Dm = (\Rm^b)^\TT\Rm^a.
\end{align}

{Up to this point, we have presented all the necessary Riemannian components, enabling us to establish the \ac{ukf} framework on Riemannian manifold to address the considered non-linear, constrained 9D tracking problem. For a more comprehensive review of the Riemannian manifold, readers are directed to~\cite{Lee2018Introduction,Boumal2023Introduction}.}

\subsection{UKF on Riemannian Manifold}
\label{sec:RieMUKF}

{Similar to its Euclidean counterpart, the \ac{ukf} on Riemannian manifold also includes two phases: time update and measurement update. The key difference is that the operations and transformations within these phases, as detailed in Section~~\ref{sec:RMRS}, are adapted to account for the curved geometry of the Riemannian space.}

\subsubsection{Time Update} 
{This phase predicts the next state and updates the corresponding covariance by propagating a set of carefully selected sigma points through the nonlinear system dynamics~$f(\cdot)$ and the defined operations~$\boxplus$ and~$\boxminus$. Further details are provided below.}

Let~$\bar{\zetav}_{n-1}$ denote the mean of the state~$\zetav_{n-1}$ at time~$n-1$, and~$\Pm_{n-1}$ denote the corresponding covariance matrix of the perturbation. We can generate $2N_\mathrm{p}+1$ sigma vectors using the operation~$\boxplus$ according to the following
\begin{equation}
\begin{aligned}
	\zv_0 &= \bar{\zetav}_{n-1},\label{eq:znew}\\
	\zv_i &= \bar{\zetav}_{n-1} \boxplus \!\Big(\!\sqrt{(N_\mathrm{p}+\Lambda)\Pm_{n-1}}\Big)_i,\ i=1,\dots,N_\mathrm{p},\\
	\zv_{i+N_\mathrm{p}} &= \bar{\zetav}_{n-1} \boxplus \!\Big(\!\sqrt{(N_\mathrm{p}+\Lambda)\Pm_{n-1}}\Big)_i,\ i=1,\dots,N_\mathrm{p},
\end{aligned}
\end{equation}
where~$N_\mathrm{p} = S+9$, the notation $(\sqrt{\Am})_i$ denotes the $i$-th column of the square root of the matrix~$\Am$, and~$\Lambda$ is a parameter controlling the dispersion of the generated sigma points.
The operation~$\boxplus$ guarantees that all generated sigma points lie on the defined manifold~$\mathbb{M}$, and thus the state statistics can propagate ahead with the SO(3) constraint. 

Based on the propagation of these sigma points, the mean of the state at the time instant~$n$ can be predicted as 
\begin{equation}\label{eq:updatezeta}
	\tilde{\zetav}_n = \mathrm{MoSP}\big( \{W_i^{(m)}\},\{f(\zv_i)\}\big),\\
\end{equation}
Here,~$\mathrm{MoSP}(\cdot)$ is a function that returns the mean of a set of points on the defined manifold. An iterative practice for the mean computation on a manifold is provided in~\cite[Tab.~3]{HERTZBERG201357}. Alternatively, such a mean on SO(3) can also be obtained by calculating the Karcher mean~\cite{Lawson2014Karcher}. 

Subsequently, the covariance matrix of the state estimate at the time instant~$n$ can be predicted using the operation~$\boxminus$ as
\begin{align}\label{eq:updatePn}
	\tilde{\Pm}_n \!=\! \sum_{i=0}^{2N_\mathrm{p}} W_i^{(c)}\big(f(\zv_i)\boxminus\tilde{\zetav}_n\big) \big(f(\zv_i)\boxminus\tilde{\zetav}_n\big)^{\TT} + \Pm,
\end{align}
where~$\Pm$ is given by~\eqref{eq:P}, and~$W_i^{(m)}$ and~$W_i^{(c)}$ are the associated weights to the corresponding sigma point, which are typically set as\footnote{The parameters~$\alpha$,~$\beta$, and~$\kappa$ need to be tuned. A popular choice is~$\alpha=0.001$,~$\beta=2$, and~$\kappa=0$.}
\begin{equation}\label{eq:W}
	\begin{aligned}
	W_0^{(m)} &= \frac{\Lambda}{N_\mathrm{p}+\Lambda}, \\ 
	W_0^{(c)} &= \frac{\Lambda}{N_\mathrm{p}+\Lambda} + (1-\alpha^2+\beta), \\ 
	W_i^{(m)} &= W_i^{(c)} = \frac{1}{2(N_\mathrm{p}+\Lambda)},\ i = 1,2,\dots,2N_\mathrm{p}, \\
	\Lambda &= \alpha^2(N_\mathrm{p}+\kappa) -  N_\mathrm{p}.
\end{aligned}
\end{equation}

\subsubsection{Measurement Update} 
{During this phase, the latest measurement is utilized to refine the predicted state and enhance the estimate. This involves several steps: measurement prediction, Kalman gain calculation, and state and covariance update. Predefined Riemannian components are integral to this process. Detailed elaborations follow.}

Based on the predicted state mean~$\tilde{\zetav}_n$ and perturbation covariance matrix~$\tilde{\Pm}_n$, we can further generate a set of sigma points to predict the first and second moments of the observation at time~$n$. Specifically, we can generate $2N_\mathrm{p}+1$ sigma vectors according to the following
\begin{equation}\label{eq:xnew}
	\begin{aligned}
	\xv_0 &= \tilde{\zetav}_{n},\\
	\xv_i &= \tilde{\zetav}_{n} \boxplus \Big(\sqrt{(N_\mathrm{p}+\Lambda)\tilde{\Pm}_{n}}\Big)_i,\ i=1,\dots,N_\mathrm{p},\\
	\xv_{i+N_\mathrm{p}} &= \tilde{\zetav}_{n} \boxplus \Big(\sqrt{(N_\mathrm{p}+\Lambda)\tilde{\Pm}_{n}}\Big)_i,\ i=1,\dots,N_\mathrm{p}.
	\end{aligned}
\end{equation}
Then, the mean vector and covariance matrix of the observation estimate at the time~$n$ can be predicted as
\begin{align}
	\tilde{\rhov}_n&\!=\!\sum_{i=0}^{2N_\mathrm{p}} W_i^{(m)}h(\xv_i),\label{eq:etatilde}\\
	\tilde{\Qm}_{n}&\!=\!\sum_{i=0}^{2N_\mathrm{p}} W_i^{(c)}\big(h(\xv_i)\!-\!\tilde{\rhov}_n\big) \big(h(\xv_i)\!-\!\tilde{\rhov}_n\big)^{\TT}\! +\! \Sigmam_n.\label{eq:Rtilde}
\end{align}
Note that here~$\Sigmam_n$ is the covariance matrix of the observation perturbation~$\Delta\rhov_n$ in~\eqref{eq:MUnew}. As demonstrated in~\eqref{eq:Jinv},~$\Sigmam_n=\Jm^{-1}(\zetav_n)$. However, the true \ac{ue} state~$\zetav_n$ is typically unavailable. To address this, we propose to approximate~$\Sigmam_n$ as follows
\begin{align}\label{eq:SigmaApprox}
	\hat{\Sigmam}_n = \epsilon \mathbf{I} + \sum_{i=0}^{2N_\mathrm{p}} W_i^{(c)}\Jm^{-1}(\xv_i),
\end{align}
where~$\epsilon$ is an regularization parameter and a smaller~$\epsilon$ represents a higher level of trust in the calculated \acp{fim}~$\Jm(\xv_i)$. Our simulation suggests that the value of~$\epsilon$ should be adjusted according to the dimension of~$\Jm(\xv_i)$, which depends on the system configuration and the visibility of the \ac{ue} to the RISs and satellites. Generally, when the dimension of~$\Jm(\xv_i)$ is higher, assigning a lower value to~$\epsilon$ tends to yield better performance. 

Next, the Kalman gain is obtained as
\begin{align}\label{eq:Kn}
	\Km_{n}=\tilde{\Qm}_{\zeta\rho,n}\tilde{\Qm}_{n}^{-1},
\end{align}
where $\tilde{\Qm}_{\zeta\rho,n}$ denotes the cross-covariance matrix between the predicted state and observation, specified as follows:
\begin{align}\label{eq:Qn}
	\tilde{\Qm}_{\zeta\rho,n}=\sum_{i=0}^{2N_\mathrm{p}} W_i^{(c)}\big(f(\zv_i)\boxminus\tilde{\zetav}_n\big) \big(h(\xv_i)-\tilde{\rhov}_n\big)^{\TT}.
\end{align}

Finally, we use the acquired observation vector~$\hat{\rhov}_n$ at the time~$n$ to correct the state estimation and covariance matrix as
\begin{align}
	\hat{\zetav}_n &= \tilde{\zetav}_n \boxplus \Km_{n}(\hat{\rhov}_n-\tilde{\rhov}_n), \label{eq:etahat}\\
	\hat{\Pm}_n &= \tilde{\Pm}_n - \Km_{n} \tilde{\Qm}_{n} \Km_{n}^\TT. \label{eq:Phat}
\end{align}

The procedure of the proposed \ac{ukf} method based on the Riemannian manifold is summarized in Algorithm~\ref{algo1}.

\begin{algorithm}[t]
  \renewcommand{\algorithmicrequire}{\textbf{Input:}}
  \renewcommand{\algorithmicensure}{\textbf{Output:}}
  \caption{9D Tracking Using Constrained UKF}
  \label{algo1}
  \begin{algorithmic}[1]
      \REQUIRE Observations~$\{\hat{\rhov}_n\}$.
      \ENSURE Hidden states estimates~$\{\hat{\zetav}_n\}$.
      \STATE Initialize with: $\hat{\zetav}_{-1}=\mathbb{E}{\zetav}_{-1}$ and $\hat{\Pm}_{-1} = \mathbf{0}_{(S+9)\times(S+9)}$.
      \FOR {$n\geq 0$}
      	\STATE  \textbf{------ \textit{Time Update:}}
      	\STATE Generate~$2N_\mathrm{p}+1$ sigma points~$\{\zv_i\}$ and the corresponding weights~$\{W_i^{(m)},W_i^{(c)}\}$ according to~\eqref{eq:znew} and~\eqref{eq:W}.
      	\STATE Obtain $\tilde{\zetav}_n$ and~$\tilde{\Pm}_n$ according to~\eqref{eq:updatezeta} and~\eqref{eq:updatePn}.
      	\STATE \textbf{------ \textit{Measurement Update:}}
      	\STATE Generate~$2N_\mathrm{p}+1$ sigma points~$\{\xv_i\}$  according to~\eqref{eq:xnew}.
      	\STATE Obtain~$\tilde{\rhov}_n$ and~$\tilde{\Qm}_{n}$ according to~\eqref{eq:etatilde},~\eqref{eq:Rtilde}, and~\eqref{eq:SigmaApprox}.
      	\STATE Compute the cross-covariance~$\tilde{\Qm}_{\zeta\rho,n}$ according to~\eqref{eq:Qn}.
      	\STATE Compute the Kalman gain~$\Km_n$ according to~\eqref{eq:Kn}.
      	\STATE Obtain the estimates of the mean and covariance matrix of the hidden state~$\hat{\zetav}_{n}$ and~$\hat{\Pm}_{n}$ according to~\eqref{eq:etahat} and~\eqref{eq:Phat}.
      \ENDFOR
  \end{algorithmic}
\end{algorithm}

\section{Numerical Results}
\label{sec:results}

\subsection{The Simulation Setup}

\begin{figure}[t]
  \centering
  \includegraphics[width=1\linewidth]{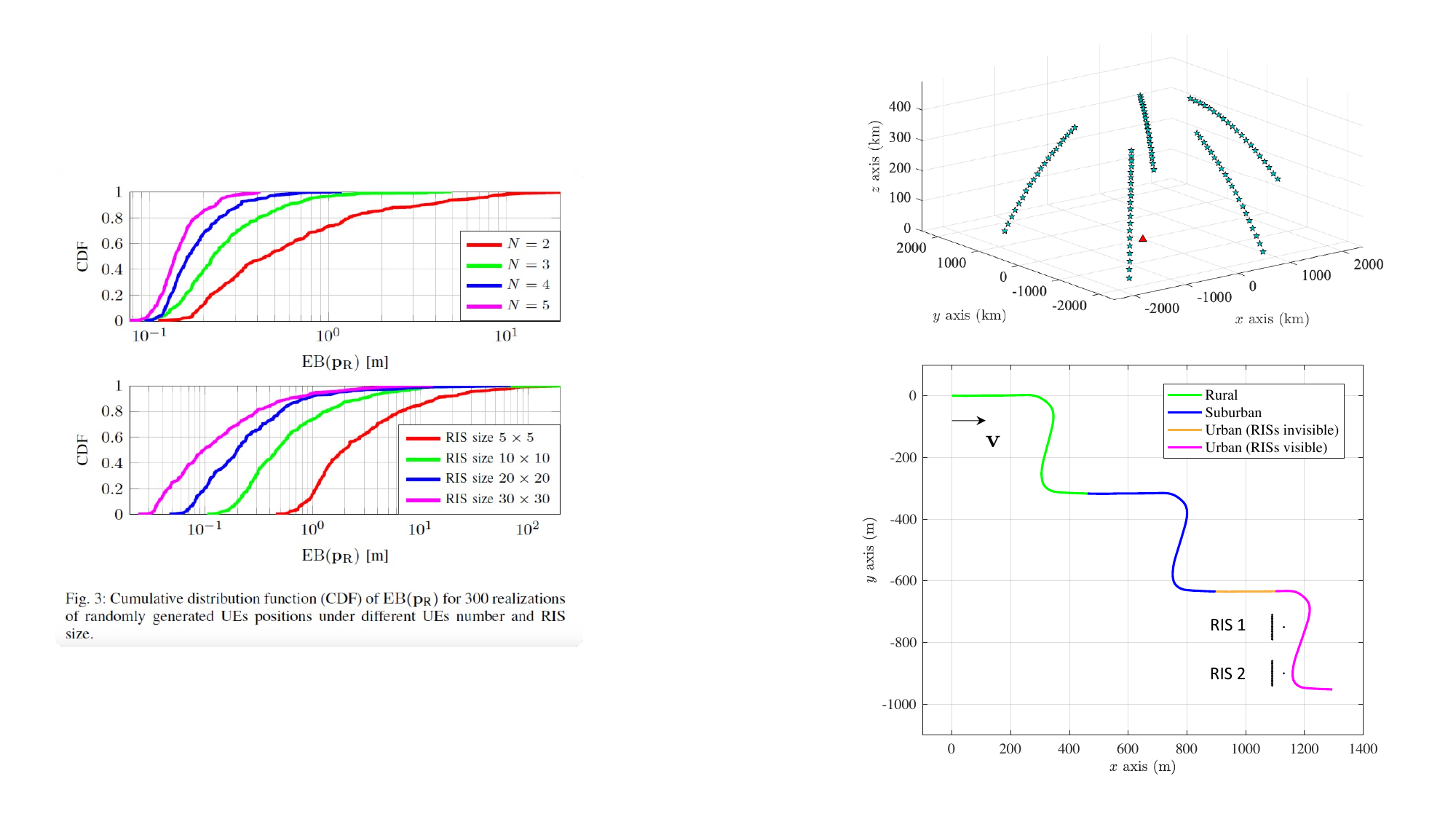}
  \caption{ 
      The trajectory of the generated 5 LEO satellites during a 180-second observation. The red triangle indicates the area of interest.
    }
  \label{fig_constellation}
\end{figure}

\begin{figure}[t]
  \centering
  \includegraphics[width=1\linewidth]{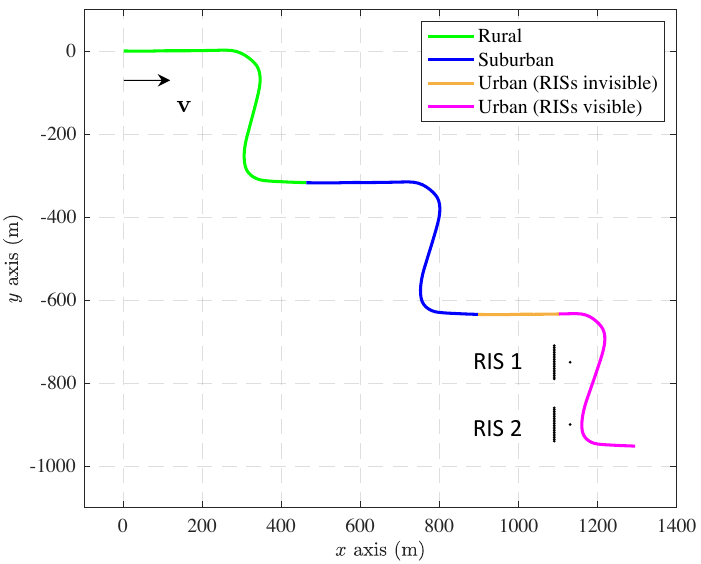}
  \caption{ 
      The demonstration of the simulated UE trajectory traversing rural, suburban, and urban regions. Two RISs are deployed in the urban area, dividing it into two segments: the visible area and the invisible area of the RISs. Here, we use two dots to indicate the normal directions of the two RISs.
    }
  \label{fig_UEstate}
\end{figure}


During the simulations, the constellation and movement of the LEO satellites with respect to the area of interest are depicted in Fig.~\ref{fig_constellation}, which are generated using the QuaDRiGa toolbox~\cite{Burkhardt2014QuaDRiGa}. Meanwhile, the trajectory of the UE is depicted in Fig.~\ref{fig_UEstate}, which is segmented into rural, suburban, and urban segments to reflect diverse environmental conditions. Different values of Rician fading factor~$K_f$ are allocated to the three environments, according to the real measured results in~\cite{Zhu2014Antenna}. In addition, two RISs are strategically deployed in the urban region, which further divides the urban area into two zones: i) the RIS-visible area and ii) the RIS-invisible area. 

Throughout the simulations, the channel gain is computed based on the large-scale path loss model provided in~\cite{3GPPSAT}. Specifically, the satellite-\ac{ue}/\ac{ris} large-scale path loss is composed of components as follows:
\begin{align}\label{eq:PL}
	\mathrm{PL} = \mathrm{FSPL} + \mathrm{SF} + \mathrm{CL} + \mathrm{PL}_g + \mathrm{PL}_s\ (\text{dB}),
\end{align}
where~$\mathrm{PL}$ is the total path loss,~$\mathrm{FSPL}$ is the free space path loss,~$\mathrm{SF}$ is the shadow fading loss represented by a random Gaussian variable,~$\mathrm{CL}$ is the clutter loss,~$\mathrm{PL}_g$ is the attenuation due to atmospheric absorption, and~$\mathrm{PL}_s$ is the attenuation due to either ionospheric or tropospheric scintillation. Typically, the clutter loss~$\mathrm{CL}=\unit[0]{dB}$ in the \ac{los} condition and the ionospheric scintillation in~$\mathrm{PL}_s$ can be ignored for frequencies above \unit[6]{GHz}~\cite{3GPPSAT}. In addition, the small-scale fading has already been characterized by the Rician model~\eqref{eq:Rician}.

\begin{figure}[t]
  \centering
  \includegraphics[width=0.96\linewidth]{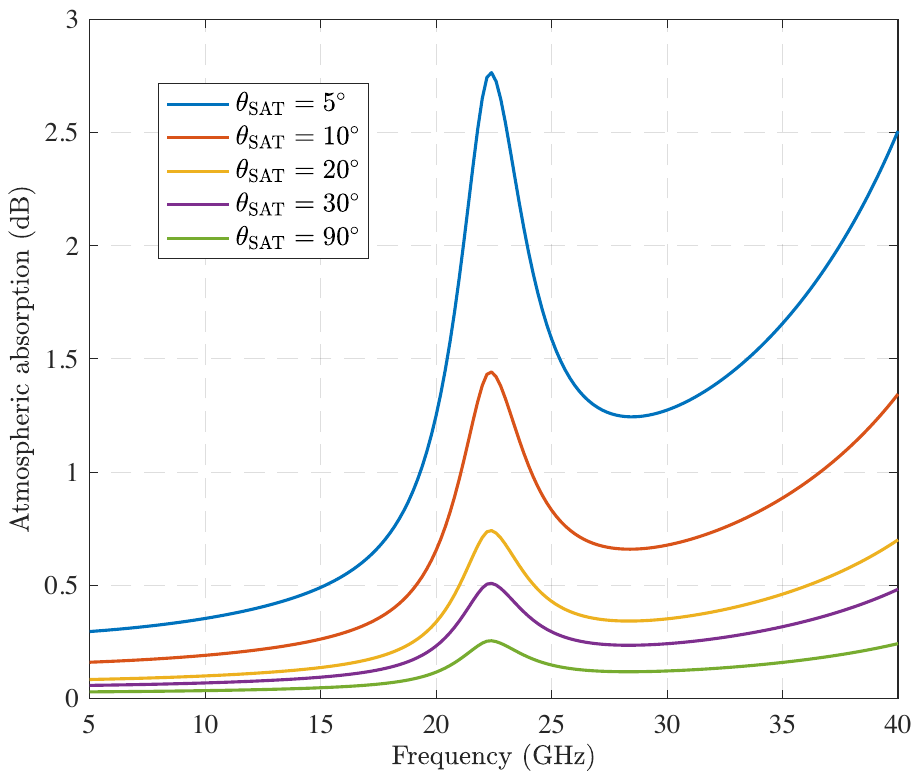}
  \caption{ 
      The evaluation of the atmospheric absorption vs. signal frequency and satellite elevation angle, based on the computation method in~\cite{Series2019Attenuation,Akram2024MATLAB}.  
    }
  \label{fig_AtmAbs}
\end{figure}

The free space path loss and shadow fading loss are calculated according to the method in~\cite{Jaeckel20225G} as
\begin{align}
	\mathrm{FSPL}(d,f_c) &= 32.45 + 20\log_{10}d + 20\log_{10}f_c~(\text{dB}), \\
	\mathrm{SF}(\theta_\mathrm{SAT}) &= X(V_\sigma + V_\theta\log_{10}\theta_\mathrm{SAT})~(\text{dB}),
\end{align}
where~$d$ represents the distance between the transmitter and receiver in meters,~$f_c$ denotes the carrier frequency in gigahertz (GHz),~$X$ is a Gaussian distributed random variable with zero-mean and unit variance,~$V_\sigma$ and~$V_\theta$ are parameters varying from environments, and~$\theta_\mathrm{SAT}$ is the elevation angle of the satellite to the \ac{ue} measured in radians. The selection of the values~$V_\sigma$ and~$V_\theta$ are listed in Table~\ref{tab_ESP}, based on the data provided in~\cite{Jaeckel20225G}. The atmospheric absorption~$\mathrm{PL}_g$ depends on the signal frequency~$f_c$ and the satellite elevation angle~$\theta_\mathrm{SAT}$. A detailed calculation method based on layered approximation has been provided by the International Telecommunication Union (ITU) in~\cite{Series2019Attenuation} and a MATLAB script is available in~\cite{Akram2024MATLAB}. As an example, Fig.~\ref{fig_AtmAbs} presents the strength of the atmospheric absorption vs. signal frequency in several sample elevation angles. Generally, a lower elevation angle~$\theta_\mathrm{SAT}$ results in a higher atmospheric absorption~$\mathrm{PL}_g$. Since the tropospheric attenuation in~$\mathrm{PL}_s$ is latitude-dependent, there is no generic analytic evaluation model. An illustrative example of the tropospheric scintillation attenuation at \unit[20]{GHz} in Toulouse, France is provided in~\cite[Table 6.6.6.2.1-1]{3GPPSAT}, which is adopted as a reference in this work. Besides, while the path loss of the link between the satellite and \ac{ue}/\ac{ris} is computed according to~\eqref{eq:PL}, the path loss of the link between the \ac{ris} and \ac{ue} is obtained by keeping~$\mathrm{FSPL}$ and~$\mathrm{SF}$ only.

To conduct a more realistic evaluation, we further take the \ac{los} probability into account. That is, we set the amplitude of the LoS channel gain as the expectation~$P(\theta_\mathrm{SAT})10^{\mathrm{-PL}/{20}}$, where~$P(\theta_\mathrm{SAT})$ denotes the \ac{los} probability. The value of~$P(\theta_\mathrm{SAT})$ varies across different environments based on the data provided in~\cite[Table 6.6.1-1]{3GPPSAT}, which is also summarized in Table~\ref{tab_ESP}. Additionally, the phases of these channel gains are randomly generated through a uniform distribution between~$0$ and~$2\pi$~\cite{Zheng2023JrCUP,Chen2024Multi,Zheng2023Coverage}.

\begin{table}[t]
  \renewcommand{\arraystretch}{1.2}
  \begin{center}
  \caption{Environment-Specific Parameters}\vspace{-0.5em}
  \label{tab_ESP}
  \begin{tabular}{  c !{\vrule width1pt} c | c | c}
    \Xhline{1pt}
    \ \ \ \ Parameter\ \ \ \  &\ \ \ \ Rural\ \ \ \  &\ Suburban\ \ &\ \ \ \ Urban\ \ \ \ \\
    \Xhline{1pt}
    $V_\sigma$ & 1.40 & 1.45 &  0.10\\
    \hline
    $V_\theta$ & 1.00 & 0.85 &  0.00\\
    \hline
    $P(10^\circ)$ & \multicolumn{2}{c|}{ \unit[78.2]{\%} } &  \unit[24.6]{\%} \\
    \hline
    $P(20^\circ)$ & \multicolumn{2}{c|}{ \unit[86.9]{\%} } &  \unit[38.6]{\%} \\
    \hline
    $P(30^\circ)$ & \multicolumn{2}{c|}{ \unit[91.9]{\%} } &  \unit[49.3]{\%} \\
    \hline
    $P(40^\circ)$ & \multicolumn{2}{c|}{ \unit[92.9]{\%} } &  \unit[61.3]{\%} \\
    \hline
    $P(50^\circ)$ & \multicolumn{2}{c|}{ \unit[93.5]{\%} } &  \unit[72.6]{\%} \\
    \hline
    $P(60^\circ)$ & \multicolumn{2}{c|}{ \unit[94.0]{\%} } &  \unit[80.5]{\%} \\
    \hline
    $P(70^\circ)$ & \multicolumn{2}{c|}{ \unit[94.9]{\%} } &  \unit[91.9]{\%} \\
    \hline
    $P(80^\circ)$ & \multicolumn{2}{c|}{ \unit[95.2]{\%} } &  \unit[96.8]{\%} \\
    \hline
    $P(90^\circ)$ & \multicolumn{2}{c|}{ \unit[99.8]{\%} } &  \unit[99.2]{\%} \\
    \Xhline{1pt}
    \end{tabular}
\end{center}
\end{table}

The covariance matrices of the acceleration and rotation measurements are defined as~$\Cm_a = \sigma_a^2\mathbf{I}_3$ and~$\Cm_\omega = \sigma_\omega^2\mathbf{I}_3$.
{We assume no prior knowledge of the channel. Therefore, all the precoders, combiners, and RIS phase shifts are randomly assigned, similar to the approach in~\cite{He2021Channel,Ozturk2023RIS}.}\footnote{{Typically, beamforming optimization for positioning and tracking in RIS-aided systems can be achieved by minimizing the estimation Cramér-Rao lower bound. However, solving this problem usually requires prior information about the channel or the UE's location. For relevant works, see, e.g.,~\cite{Fascista2022RIS,Chen2024Multi}. Nonetheless, in the tracking problem under consideration, prior information about the current \ac{ue} location can be predicted based on updates from previous time slots. Thus, the optimization of the precoders, combiners, and RIS coefficients becomes feasible. Due to space limitations, we leave this as a direction for future exploration.}} Since it has been verified that active RISs can mitigate the multiplicative fading effect and can outperform passive RISs in terms of user positioning within the practical power supplies~\cite{Zheng2023JrCUP}, we choose to deploy active RISs in our system during the following performance evaluation. We assume a power supply~$P_\mathrm{R}$ to each active RIS, and the amplification coefficient of the RIS can be calculated following~\cite[Eq.~(85)]{Zheng2023JrCUP}. The clock bias between each satellite and the \ac{ue} are randomly set between~\unit[80]{ns} and~\unit[120]{ns}, which is assumed fixed over the observation duration. Furthermore, we assume that all the LEO satellites keep the same transmission power~$P_\mathrm{T}$. Other default simulation parameters are summarized in Table~\ref{tab1}.
	
\begin{table}[t]
  \renewcommand{\arraystretch}{1.2}
  \begin{center}
  \caption{Default Simulation Parameters}\vspace{-0.5em}
  \label{tab1}
  \begin{tabular}{  c !{\vrule width1pt} c }
    \Xhline{1pt}
    \textbf{Parameter} & \textbf{Value}\\
    \Xhline{1pt}
    Number of LEO satellites~$S$ & 5 \\
    \hline
    Number of RISs~$R$ & 2 \\
    \hline  
    Number of Subcarriers $K$ & $3000$ \\
    \hline
    Number of Transmissions $G$ & $32$ \\
    \hline
    Carrier Frequency $f_c$ & $\unit[12.7]{GHz}$ \\
    \hline
    Bandwidth $B$ & $\unit[240]{MHz}$ \\
    \hline 
    Transmission Power $P_\text{T}$ & $\unit[50]{dBm}$ \\
    \hline
    Active RIS Power $P_\text{R}$ & $\unit[0]{dBm}$ \\
    \hline
    Noise PSD of Receiver \& RIS & $\unit[-174]{dBm/Hz}$ \\
    \hline
    Noise Figure of Receiver \& \ac{ris} & $\unit[0]{dB}$ \\
    \hline
    Array Size of satellite / \ac{ris} / \ac{ue}  & $4\times 4$ / $20\times 20$ / $4\times 4$\\
    \hline
    Array Spacing of satellite / \ac{ris} / \ac{ue}  & $\unit[5]{mm}$ / $\unit[5]{mm}$ / $\unit[5]{mm}$\\
    \hline
    Rural / Suburban / Urban $K_f$~\cite{Zhu2014Antenna}  & $\unit[3]{dB}$ / $\unit[2.6]{dB}$ / $\unit[1.85]{dB}$  \\
    \hline
    Acceleration standard deviation $\sigma_a$ & $\unit[0.2]{m/s^2}$\\
    \hline
    Rotation standard deviation $\sigma_\omega$ & $\unit[2]{^\circ}$\\
    \hline
    Update Interval $T_\Ut$ & \unit[1]{s} \\
    \hline
    Orbit Height of LEO satellites & \unit[500]{km} \\
    \Xhline{1pt}
    \end{tabular}
\end{center}
\end{table}

\subsection{9D Tracking Performance Evaluation}

\subsubsection{The Evaluation of the Proposed Belief Assignment Principle}

\begin{figure}[t]
  \centering
  \includegraphics[width=1\linewidth]{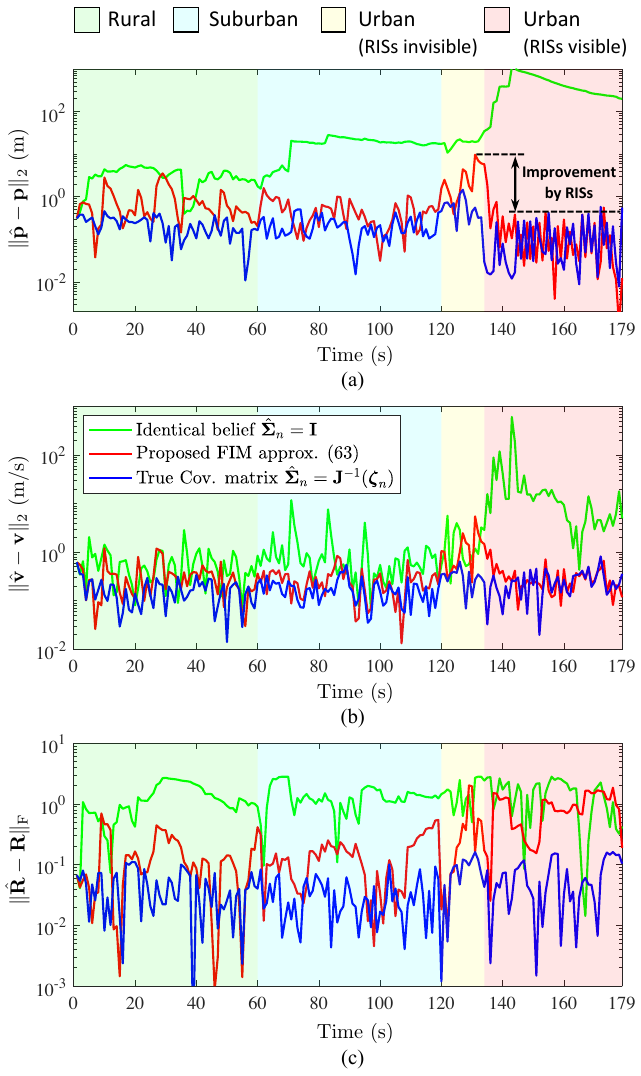}
  \caption{ 
      The comparison of the estimation errors vs. time using different assignment principles of the belief~$\hat{\Sigmam}_n$. (a) Estimation errors of the \ac{ue} position~$\pv$. (b) Estimation errors of the \ac{ue} velocity~$\mathbf{v}$. (c) Estimation errors of the \ac{ue} orientation~$\Rm$.
    }
  \label{fig_err_belief}
\end{figure}

First, we evaluate the proposed belief assignment principle as formulated in~\eqref{eq:SigmaApprox}. Fig.~\ref{fig_err_belief} presents the estimation errors of the \ac{ue} position, velocity, and orientation during a 180-second tracking duration. In this simulation, we set~$\epsilon=0.5$ in the rural, suburban, and RIS-invisible urban regions, while set~$\epsilon=0$ in the RIS-visible urban areas. As benchmarks, we compared the proposed \ac{fim}-based belief approximation~\eqref{eq:SigmaApprox} with i) the identical belief~$\hat{\Sigmam}_n=\mathbf{I}$ and ii) the true covariance matrix~$\hat{\Sigmam}_n=\Jm^{-1}(\zetav_n)$. Note that the true covariance matrix~$\Jm^{-1}(\zetav_n)$ is typically unavailable in practice, and here we adopt it as a performance limit. It is clearly shown that while the true covariance matrix belief (blue curve) yields the best performance, the proposed belief assignment principle~\eqref{eq:SigmaApprox} (red curve) significantly outperforms the identical belief assignment (green curve). In particular, we can observe that the~\ac{fim} approximation~\eqref{eq:SigmaApprox} can achieve performance comparable to that of the true covariance matrix in some regions, demonstrating the superiority of the proposed principle. 

Another important observation is that the introduction of RISs can substantially improve the accuracy of the \ac{ue} position estimation in the urban environment despite the lower Rician factor and channel gains, as highlighted in Fig.~\ref{fig_err_belief}-(a). Meanwhile, it is also revealed that the inclusion of RISs can only be positive when an appropriate belief~$\hat{\Sigmam}_n$ is assigned. This can be observed from the identical belief (green curve) in Fig.~\ref{fig_err_belief}-(a), where the tracking performance is even degraded when the RISs are introduced. In contrast, both the proposed \ac{fim} approximation and the true covariance matrix can make good use of the additional information introduced by RISs and gain substantial performance improvement for up to more than an order of magnitude. This reveals the potential of RISs in enhancing the tracking services based on LEO satellite signals. Such an improvement appears because the RISs can not only increase the received power of the satellite signals but also provide additional location references that help determine the \ac{ue}'s position~\cite{Zheng2023JrCUP,Chen2024Multi,Zheng2023Misspecified}. When it comes to the velocity and orientation estimation, however, the introduction of RISs cannot enhance the estimation performance as effectively as the position estimation, see Fig.~\ref{fig_err_belief}-(b) and Fig.~\ref{fig_err_belief}-(c). Such a phenomenon can be explained by the geometry model in Section~\ref{sec_GM}. All the channel parameters related to the RIS reflection channels contain certain information about the \ac{ue} position since all of them are the functions of~$\pv$. On the contrary, only part of these channel parameters are the functions of~$\mathbf{v}$ or~$\Rm$. Therefore, introducing RISs benefits the \ac{ue} position estimation most, while boosting the \ac{ue} velocity and orientation estimation less. 

\subsubsection{Euclidean Space vs. Riemannian Manifold}

\begin{figure}[t]
  \centering
  \includegraphics[width=1\linewidth]{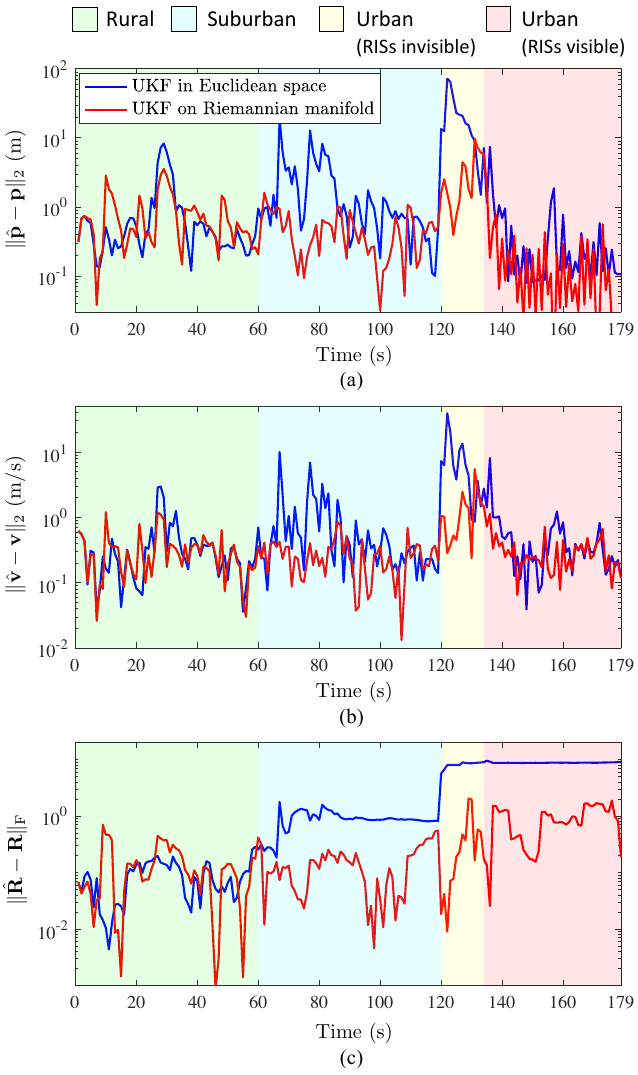}
  \caption{ 
      The comparison of the estimation errors vs. time using the \acp{ukf} in Euclidean space and on Riemannian manifold. (a) Estimation errors of the \ac{ue} position~$\pv$. (b) Estimation errors of the \ac{ue} velocity~$\mathbf{v}$. (c) Estimation errors of the \ac{ue} orientation~$\Rm$.    
    }
  \label{fig_err_manifold}
\end{figure}

Next, we examine the effectiveness of the proposed Riemannian manifold-based framework in enhancing tracking performance benchmarked against conventional Euclidean techniques.
Fig.~\ref{fig_err_manifold} compares the estimation errors of the classical \ac{ukf} in Euclidean space and the proposed \ac{ukf} on Riemannian manifold. In this simulation, both methods utilize the proposed belief assignment~\eqref{eq:SigmaApprox}. From Fig.~\ref{fig_err_manifold}, insights are obtained as follows:
\begin{itemize}
	\item In general, without the presence of RISs, the tracking performance of the system exhibits superior performance in rural regions, deteriorates in suburban areas, and experiences the poorest performance in urban environments. This phenomenon arises due to the decreased Rician factor and channel gain in closer proximity to the urban environment, thereby diminishing overall system performance. However, upon the integration of RIS, similar to the results in Fig.~\ref{fig_err_belief}, significant improvements are observed in the estimation of the \ac{ue} position and velocity (particularly position) under both methodologies. This again confirms that the involvement of RISs can potentially mitigate the negative impact of urban attenuation. Conversely, the introduction of RISs contributes minimally to the estimation of the \ac{ue} orientation. 
	\item By comparing the \ac{ukf} in Euclidean space and on Riemannian manifold, we observe that the two methods showcase a similar level of accuracy in the urban region with strong signal reception. Nonetheless, when it comes to the suburban and urban environments with weaker signal reception, the proposed Riemannian manifold-based \ac{ukf} can consistently outperform the classical Euclidean \ac{ukf}. Such superiority is particularly significant in the orientation estimation, verifying that the proposed \ac{ukf} on the Riemannian manifold can effectively preserve the SO(3) constraint and thus enhance the orientation estimation performance, which in turn benefits the position and velocity estimation as well.
\end{itemize}
 
\subsubsection{Comparison with Benchmark Methods}

\begin{figure*}[t]
  \centering
  \includegraphics[width=0.95\linewidth]{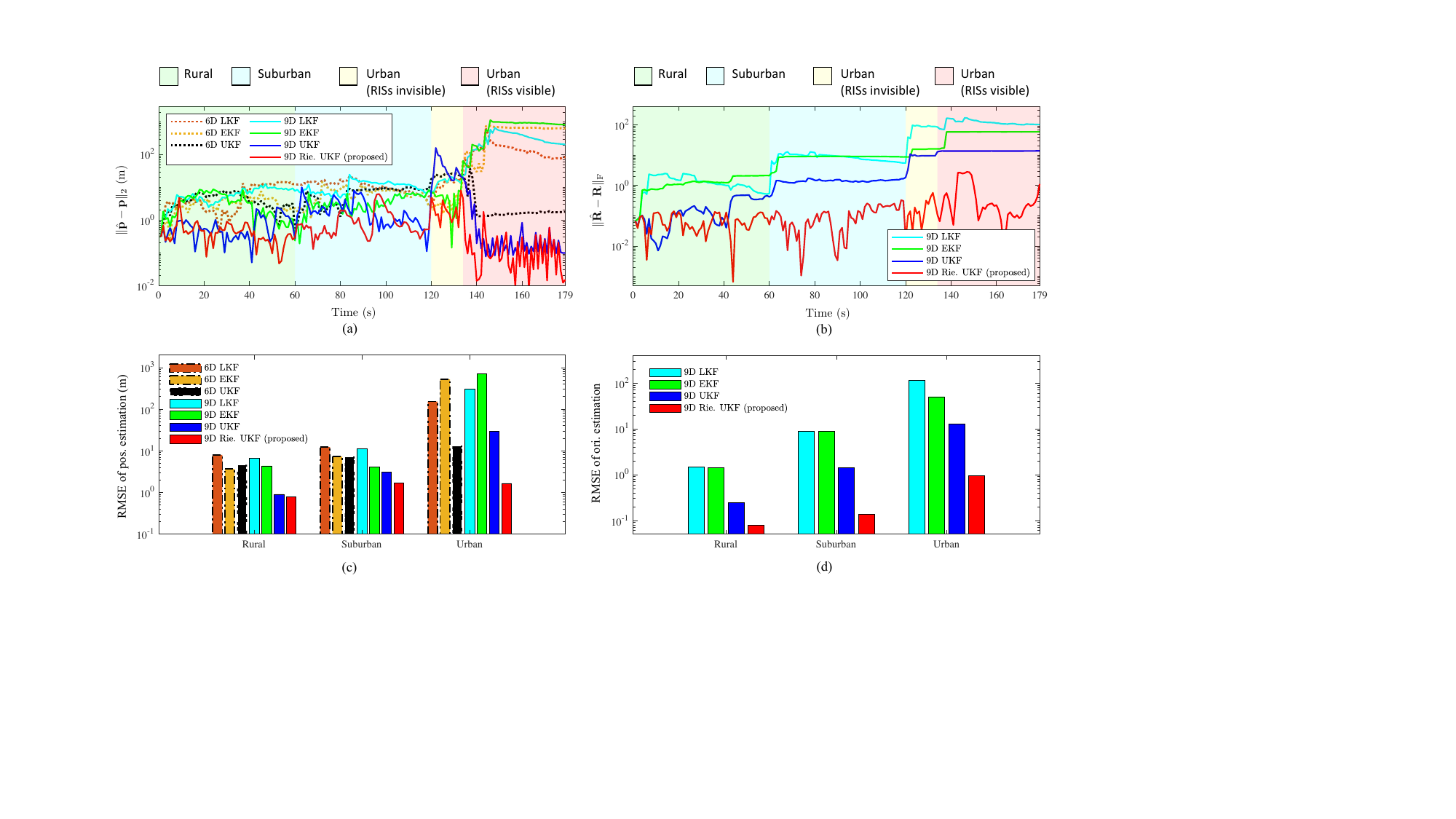}
  \caption{ 
      {The comparison of the UE position and orientation estimation using the proposed 9D Riemannian UKF and other existing methods, namely, 6D LKF, 6D EKF, 6D Euclidean UKF, 9D LKF, 9D EKF, and 9D Euclidean UKF. (a) UE position estimation error vs. time. (b) UE orientation estimation error vs. time. (c) \Ac{rmse} of UE position estimation errors in different environments. (d) \Ac{rmse} of UE orientation estimation errors in different environments. } 
    }
  \label{fig_err_BMs}
\end{figure*}

\begin{table*}[t]
  \renewcommand{\arraystretch}{1.3}
  \begin{center}
  \caption{{The Computational Complexity Evaluation of Different Kalman Filters}}\vspace{-0.5em}
  \label{tab_CC}
  \begin{tabular}{  c !{\vrule width1pt} c !{\vrule} c !{\vrule} c }
    \Xhline{1pt}
    Step & {\makecell[c]{Computation of sigma points / \\linearized coefficient matrices}} & Obs. belief approximation~\eqref{eq:SigmaApprox} &  Standard Kalman filtering process \\
    \Xhline{1pt}
    LKF & \multirow{2}{*}{$\mathcal{O}(RS^2)$} & \multirow{2}{*}{$\mathcal{O}(1)$} & \multirow{4}{*}{$\mathcal{O}(R^2S^3)+\mathcal{O}(R^2S^2\log(RS))$} \\
    \cline{1-1} 
    EKF &  &  &  \\
    \cline{1-3}  
    Euclidean UKF & \multirow{2}{*}{$\mathcal{O}(S^3)$} & \multirow{2}{*}{$\mathcal{O}(R^2S^3\log(RS))$} &  \\
    \cline{1-1} 
    Riemannian UKF &  &  &  \\
    \Xhline{1pt}
    \end{tabular}
    \begin{tablenotes}
     \item{\footnotesize{{$\bullet$ These evaluations are consistent for both the 6D and 9D implements. Again, here,~$S$ is the number of satellites while~$R$ denotes the number of RISs. In addition to the listed steps, these Kalman filters still need to compute the time and measurement update functions~$f$ and~$h$ at each iteration, whose computational complexities are given by~$\mathcal{O}(S)$ and~$\mathcal{O}(RS^2)$, respectively. Compared to the classical Euclidean UKF, the extra complexity introduced by the defined projection~$\boxplus$ and~$\boxminus$ in the proposed Riemannian UKF is on the order of~$\mathcal{O}(S)$.}}} 
   \end{tablenotes}
\end{center}
\end{table*}

Now we compare the proposed 9D Riemannian UKF with other existing methods to affirm its superiority. Fig.~\ref{fig_err_BMs} evaluates the tracking performance of the proposed method and benchmark methods. Since the UE velocity estimation typically exhibits a similar trend to the position (as shown in Fig.~\ref{fig_err_belief} and~\ref{fig_err_manifold}), we only present the estimation errors of UE position and orientation in Fig.~\ref{fig_err_BMs} for compactness. We compare our method with the linearized Kalman filter (LKF), the extended Kalman filter (EKF), and the classical \ac{ukf} in Euclidean space, which are widely adopted techniques for nonlinear filtering~\cite{Sayed2022Inference}. Additionally, we note that there are numerous works that aim to address 6D (3D position and 3D velocity) tracking without orientation estimation capability~\cite{Honghui2002Direct,Chen2012Kalman}. To provide a comprehensive comparison, we also perform such estimation in the 6D Euclidean space for LKF, EKF, and UKF, as shown in Fig.~\ref{fig_err_BMs}. Note that in such 6D estimations, the UE orientation-related channel parameters such as \acp{aoa} at the UE become nuisance parameters. When evaluating the UKF (both the classical Euclidean UKF and the proposed Riemannian UKF), we utilize the proposed FIM approximation~\eqref{eq:SigmaApprox} to assign the observation belief~$\hat{\bm{\Sigma}}_n$. For LKF and EKF, we use the identity belief since they do not compute sigma points, making~\eqref{eq:SigmaApprox} unavailable.

As demonstrated in Fig.~\ref{fig_err_BMs}, our proposed method outperforms all the adopted benchmark methods in both UE position and orientation estimations. This superiority occurs consistently across different environments, confirming its effectiveness and robustness. Additionally, we observe that the UKF generally outperforms the LKF and EKF, indicating that the observation functions maintain strong non-linearity, thus making the zeroth and first-order Taylor approximations less effective than the sigma point propagation. Furthermore, we analyze and summarize the computational complexities of these Kalman filters in Table~\ref{tab_CC}. In conclusion, two main insights are obtained: (i) The UKF outperforms the LKF and EKF at the expense of higher computational complexity, which is mainly introduced by the proposed belief assignment principle~\eqref{eq:SigmaApprox}; 
(ii) The overall computational complexity of the proposed Riemannian UKF is on par with that of the classical Euclidean UKF, as the additional complexity introduced is only of the order $\mathcal{O}(S)$.
	
\subsection{The Impact of the State and Observation Uncertainty}

Now we investigate how the state and observation uncertainty affect the tracking performance. As clarified in Section~\ref{sec:UR}, the uncertainty in the state arises from measurement errors in \ac{ue}'s acceleration and rotation, which can be represented by the standard deviation~$\sigma_a$ and~$\sigma_\omega$. On the other hand, the uncertainty on the observation is introduced by the estimation errors of the channel parameters, which can be controlled by the number of subcarriers~$K$~\cite{Chen2022A}. Typically, a greater number of subcarriers results in a higher \ac{fim}, thereby reducing the observation uncertainty in channel parameters. For compactness, here we only show the estimation performance of \ac{ue} position and orientation while omitting the evaluation of \ac{ue} velocity, due to the perturbation in the \ac{ue} velocity is injected in the same pattern as that in the \ac{ue} position, as demonstrated in~\eqref{eq:f}. The behavior of the \ac{ue} velocity is the same as the position.

\begin{figure}[t]
  \centering
  \includegraphics[width=\linewidth]{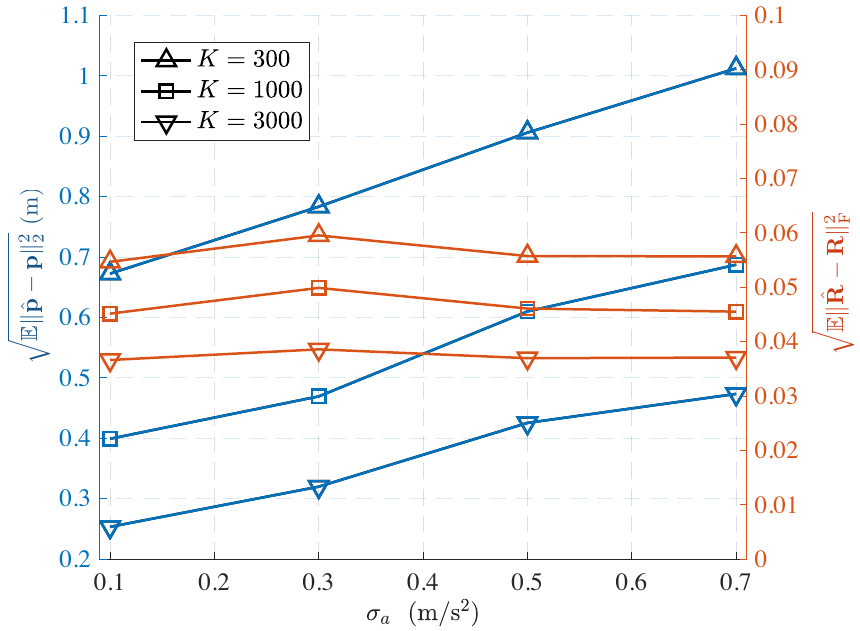}
  \caption{ 
      The evaluation of the estimation errors in the \ac{ue} position and orientation vs. the standard deviation~$\sigma_a$ of the \ac{ue} acceleration measurement. The tests are conducted in different cases that~$K=\{300,1000,3000\}$.
    }
  \label{fig_sigmaa}
\end{figure}

\begin{figure}[t]
  \centering
  \includegraphics[width=\linewidth]{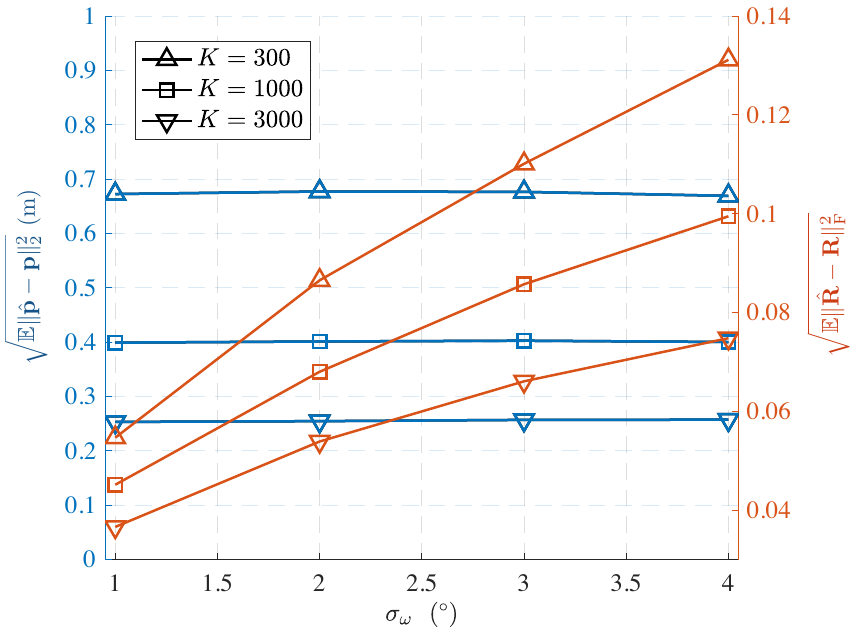}
  \caption{ 
      The evaluation of the estimation errors in the \ac{ue} position and orientation vs. the standard deviation~$\sigma_\omega$ of the \ac{ue} rotation measurement. The tests are conducted in different cases that~$K=\{300,1000,3000\}$. 
    }
  \label{fig_sigmaw}
\end{figure}

\subsubsection{The Impact of~$\sigma_a$}
We first evaluate the estimation \acp{rmse} of the \ac{ue} position and orientation vs. the standard deviation~$\sigma_a$ of the \ac{ue} acceleration measurement. In this evaluation, we compute the \ac{rmse} using the proposed Riemannian manifold-based \ac{ukf} over $\sigma_a=\unit[\{0.1,0.3,0.5,0.7\}]{m/s^2}$ for the cases: (i)~$K=300$; (ii)~$K=1000$; (iii)~$K=3000$. During this simulation, the standard deviation of the rotation measurement is fixed as~$\sigma_\omega = \unit[1]{^\circ}$. The results are presented in Fig.~\ref{fig_sigmaa}. It is observed that the estimation \ac{rmse} of the \ac{ue} position increases with an increase in~$\sigma_a$, while that of the \ac{ue} orientation keeps flat over different~$\sigma_a$ values. This phenomenon showcases the robustness of the proposed method in the orientation estimation to the acceleration measurement errors. Apart from that, it is observed that increasing the number of subcarriers~$K$ (i.e., reducing the uncertainty in the observation) can also effectively improve the estimation accuracy.

\subsubsection{The Impact of~$\sigma_\omega$}
Then we evaluate the estimation \acp{rmse} of the \ac{ue} position and orientation vs. the standard deviation~$\sigma_\omega=\{1^\circ,2^\circ,3^\circ,4^\circ\}$ of the \ac{ue} rotation measurement. The simulation is carried out in the same setup as in Fig.~\ref{fig_sigmaa} while the acceleration perturbation is fixed as~$\sigma_a=\unit[0.1]{m/s^2}$. The result is depicted in Fig.~\ref{fig_sigmaw}. Opposite to the results in Fig.~\ref{fig_sigmaa}, we can observe from Fig.~\ref{fig_sigmaw} that the estimation \ac{rmse} of the \ac{ue} orientation increases with the increase of~$\sigma_\omega$, while the estimation \ac{rmse} of the \ac{ue} position keeps flat over different~$\sigma_\omega$ values. This indicates the robustness of the proposed method in the position estimation to the rotation measurement errors. In addition, the same phenomenon is observed that enlarging~$K$ can help improve tracking performance. 

\subsubsection{Observation Uncertainty vs. Transmission Overhead}

\begin{figure}[t]
  \centering
  \includegraphics[width=\linewidth]{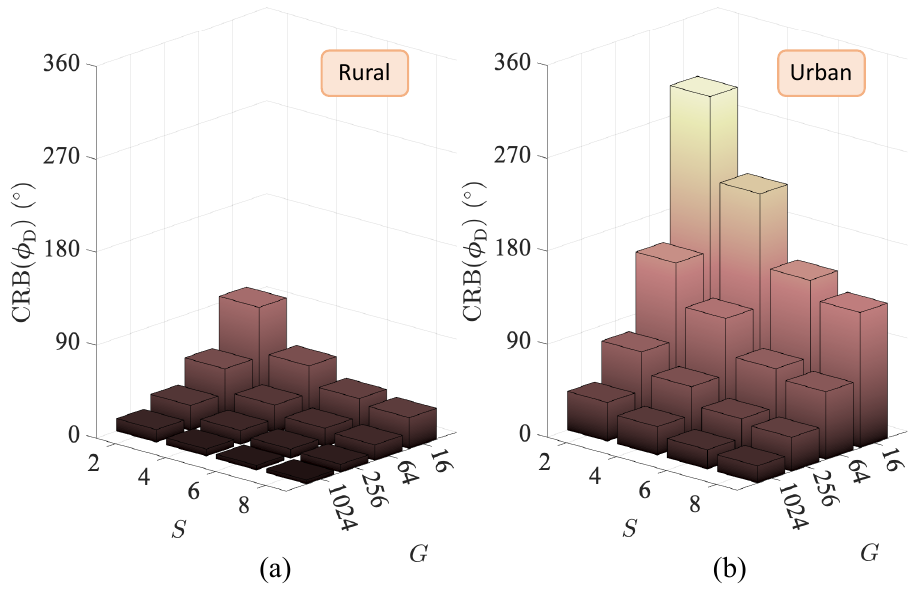}
  \caption{ 
      The observation uncertainty vs. the transmission overhead. The observation uncertainty is represented by the \ac{crb} of~$\phiv_\mathrm{D}$ while the transmission overhead is determined by the number of transmissions~$G$ and the number of satellites~$S$. The evaluation is conducted under the rural and urban scenarios, respectively.  
    }
  \label{fig_CRB}
\end{figure}

The results shown above indicate that reducing the uncertainty in the observations (by increasing~$K$) can consistently improve the estimation performance for all unknowns. Besides the number of subcarriers~$K$, the uncertainty in the observations can also be lowered by increasing the number of transmissions~$G$ and the number of satellites~$S$. However, increasing~$G$ or~$S$ can also notably increase the transmission overhead. As demonstrated in Fig.~\ref{fig_framework}, the total overhead during a single update interval~$T_\Ut$ is~$GS$. Therefore, we now evaluate the observation uncertainty vs. transmission overhead. Specifically, we select a time instant~$n$ in the RIS-visible area and compute the \ac{crb} of the \acp{aod}~$\phiv_\mathrm{D}$ from the RISs to the \ac{ue} as a representative of the uncertainty in the observation, which is given by
\begin{equation}
	\text{CRB}(\phiv_\mathrm{D}) = \sqrt{\text{Tr}([\Jm^{-1}(\zetav_n)]_{R+1:3R,R+1:3R})}.
\end{equation}

Fig.~\ref{fig_CRB} presents the evaluation of~$\text{CRB}(\phiv_\mathrm{D})$ vs.~$\{G,S\}$, where both the rural and urban scenarios are assigned and tested. It is clearly shown that there exists a trade-off between the transmission overhead and the observation uncertainty. Generally, increasing the number of transmissions or satellites can always help reduce the uncertainty in the channel parameter observations. Comparing the results in the rural and urban environments, we can further conclude that the urban environment would also increase the uncertainty in the channel parameters due to the higher signal attenuation. Hence, it is suggested that a higher transmission overhead is required in urban regions to combat this uncertainty elevation.

\subsection{The Impact of the System Configuration}

\begin{figure}[t]
  \centering
  \includegraphics[width=1\linewidth]{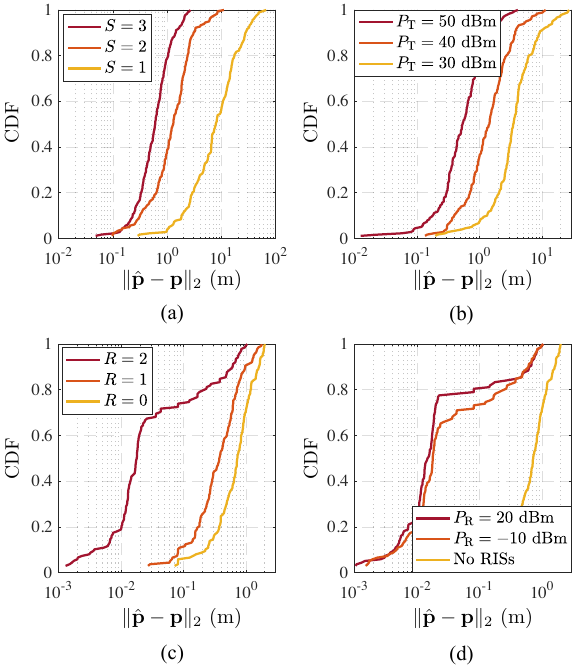}
  \caption{ 
      The CDF evaluation of the estimation \acp{rmse} of the \ac{ue} position under different system setups. (a)~The number of satellites~$S=\{1,2,3\}$. (b)~The satellite transmission power~$P_\mathrm{T}=\{\unit[30]{dBm},\unit[40]{dBm},\unit[50]{dBm}\}$. (c)~The number of RISs~$R=\{0,1,2\}$. (d)~The power supply of RISs~$P_\mathrm{R}=\{\unit[-10]{dBm},\unit[20]{dBm}\}$ and the case without RISs. 
    }
  \label{fig_para}
\end{figure}

Finally, Fig.~\ref{fig_para} illustrates the \ac{cdf} of the estimation error of the \ac{ue} position over different system setups. These results demonstrate the impact of the system parameters including the number of LEO satellites~$S$, the transmission power at LEO satellites~$P_\mathrm{T}$, the number of \acp{ris}~$R$, and the power supply of RISs~$P_\mathrm{R}$. To better illustrate the impact of these parameters, the \acp{cdf} over different RIS numbers and powers (i.e., Fig.~\ref{fig_para}-(c) and Fig.~\ref{fig_para}-(d)) are collected from the RISs-visible urban region only. It is observed that increasing the number of satellites or the transmission power of satellites can effectively enhance the overall tracking performance. On the other hand, by adding more RISs or elevating the RIS power supply, the tracking performance can also be improved. This indicates that both LEO satellites and RISs contribute to improving tracking performance and the collaboration between them showcases a promising potential for advanced user tracking solutions across complex environments.

\section{Conclusion}
\label{sec:conclusion}

This paper studies the 9D user tracking problem within a hybrid terrestrial and non-terrestrial wireless system, integrating LEO satellites and RISs. A novel Riemannian manifold-based UKF method is proposed to simultaneously track the 3D position, 3D velocity, and 3D orientation of a ground user, effectively handling challenges posed by nonlinear observation functions, constrained UE states, and unknown observation statistics. Numerical simulations demonstrate the effectiveness and robustness of the proposed tracking method. Meanwhile, it is also revealed that the inclusion of RISs can substantially enhance the tracking performance. Given the potential of such hybrid networks, future research could focus on optimizing collaborative beamforming design, developing effective protocols, and exploring the multipath effect.

\appendices
\section{Calculation of \ac{fim}~$\Jm$}\label{appen:Jch}
\setcounter{equation}{0}

In addition to~$\rhov$ defined in~\eqref{eq:eta}, for each satellite, we can further collect the other unknown nuisance parameters as
\begin{align}
	&\xiv_s = \big[ \mathfrak{R}(\alpha_s^0),\mathfrak{I}(\alpha_s^0),\mathfrak{R}(\alphav_{\mathrm{r}}^0),\mathfrak{I}(\alphav_{\mathrm{r}}^0)\big]^\TT\!\in\!\mathbb{R}^{2(R+1)},
\end{align}
where~$\alphav_{\mathrm{r}}^0\in\mathbb{C}^R$ is a vector collecting~$\alpha_r^0,\ r=1,2,\dots,R$. 
Furthermore, we stack the noise-free version of received signals from the~$s$-th satellite over~$G$ transmissions and~$K$ subcarriers as
\begin{multline}
	\bm{\ell}_s = [\ell_s^{1}(1),\dots,\ell_s^{K}(1),\dots,\ell_s^{1}(g),\dots,\ell_s^{K}(g),\dots,\\
	\ell_s^{1}(G),\dots,\ell_s^{K}(G)]^\TT\in\mathbb{C}^{GK}.
\end{multline}
By assuming the noise at the signal across different transmissions, subcarriers, and satellites are independent, we can define the following covariance matrix of the received signals
\begin{equation}
	\Cm = \text{blkdiag}(\Cm_1,\dots,\Cm_s,\dots,\Cm_S),
\end{equation}
where~$\Cm_s=\text{diag}([C_s^{1}(1),\dots,C_s^{K}(1),\dots,C_s^{1}(g),\dots,C_s^{K}(g),\\ \dots, C_s^{1}(G),\dots,C_s^{K}(G)]^\TT)$.
Then, according to the Slepian-Bangs formula~\cite{Kay1993Fundamentals}, the \ac{fim} of the channel parameters can be calculated as
\begin{equation}
	\Jm_\mathrm{ch} = 2\mathfrak{R}(\Dm^\HH\Cm^{-1}\Dm),
\end{equation}
where the Jacobian matrix~$\Dm$ is defined as
\begin{equation}
	\Dm = \begin{bmatrix}
		\frac{\partial \bm{\ell}_1}{\partial \rhov_0} & \frac{\partial \bm{\ell}_1}{\partial \rhov_1} & \cdots & \mathbf{0} & \frac{\partial \bm{\ell}_1}{\partial \xiv_1} & \cdots & \mathbf{0}\\
		\vdots & \vdots & \ddots & \vdots & \vdots & \ddots & \vdots\\
		\frac{\partial \bm{\ell}_S}{\partial \rhov_0} & \mathbf{0} & \cdots & \frac{\partial \bm{\ell}_S}{\partial \rhov_S} & \mathbf{0} & \cdots & \frac{\partial \bm{\ell}_S}{\partial \xiv_S}
	\end{bmatrix}.
\end{equation}

Since we do not extract any information from nuisance parameters vectors~$\xiv_s, s=1,\dots,S$, the corresponding contribution of these parameters to \ac{fim}~${\Jm}_\mathrm{ch}$ needs to be removed. To this end, we partition~${\Jm}_\mathrm{ch}$ as 
\begin{equation}
	{\Jm}_\mathrm{ch} = \begin{bmatrix}
		\Xm & \Ym \\
		\Ym^\TT & \Zm
	\end{bmatrix},
\end{equation}
where the dimension of~$\Xm$ is~$\big(5R+(R+6)S\big)\times\big(5R+(R+6)S\big)$, and the dimension of~$\Zm$ is~$2S(R+1)\times 2S(R+1)$. Then the \ac{fim} of channel parameters without nuisance parameters can be calculated as
\begin{equation}
	{\Jm} = \Xm - \Ym\Zm^{-1}\Ym^\TT.
\end{equation}

\bibliography{references}
\bibliographystyle{IEEEtran}

\end{document}